\newif\ifarxiv
\arxivtrue

\documentclass[pagebackref,11pt]{article}
\usepackage[a4paper,margin=3cm]{geometry}

\usepackage{lineno}
\ifarxiv{}\else{}
\linenumbers
\fi{}

\usepackage[utf8]{inputenc}
\usepackage[T1]{fontenc}
\usepackage[x11names, table]{xcolor}
\usepackage{amssymb,amsmath}
\usepackage[outline]{contour}
\usepackage{enumerate}
\usepackage{dsfont}
\usepackage{graphicx}
\usepackage{booktabs}
\usepackage{tabularx}
\usepackage{multicol}
\usepackage[ruled,vlined,linesnumbered]{algorithm2e}
\usepackage[numbers,sort&compress]{natbib} 
\usepackage[linkcolor=red!50!black,citecolor=green!40!black,colorlinks,unicode]{hyperref}
\usepackage[capitalize,nameinlink]{cleveref}
\usepackage{doi}
\usepackage{paralist}
\usepackage{tikz}
\usepackage{authblk}
\usetikzlibrary{calc,positioning}

\usepackage{mathtools}

\usepackage[backgroundcolor=gray!10,textsize=footnotesize]{todonotes}

\usepackage{multirow}

\usepackage{xargs} 
\newcommandx{\set}[2][1=1]{\ensuremath{\{#1,\ldots,#2\}}}
\newcommandx{\tlog}[3][1=,3=]{\log_{#1}^{#3}(#2)}

\usepackage{amsthm}
\usepackage{thmtools}
\theoremstyle{plain}
\newtheorem{theorem}{Theorem}
\newtheorem{lemma}{Lemma}
\newtheorem{corollary}[theorem]{Corollary}
\newtheorem{observation}[lemma]{Observation}
\newtheorem{proposition}[lemma]{Proposition}
\newtheorem{rrule}{Reduction Rule}
\theoremstyle{definition}

\newtheorem{problem}{Problem}
\declaretheorem[style=definition,name=Construction,qed=$\diamond$]{construction}
\theoremstyle{remark}
\newtheorem*{remark}{Remark}
 
\crefname{observation}{Observation}{Observations}
\crefname{rrule}{Reduction Rule}{Reduction Rules}
\crefname{construction}{Construction}{Constructions}
\crefname{proposition}{Proposition}{Propositions}
\crefname{theorem}{Theorem}{Theorems}
\crefname{corollary}{Corollary}{Corollaries}
\crefname{line}{Line}{Lines}
\crefname{problem}{Problem}{Problems}
\crefname{figure}{Figure}{Figures}

\Crefname{subsection}{Sec.}{Sects.}
\Crefname{section}{Sec.}{Sects.}
\Crefname{problem}{Prob.}{Probs.}
\Crefname{observation}{Obs.}{Obs.}
\Crefname{proposition}{Prop.}{Props.}
\Crefname{corollary}{Cor.}{Cors.}
\Crefname{theorem}{Thm.}{Thm.}

\newcommandx{\decprob}[6][3=Input,5=Question]{\begin{samepage}
  \begingroup
  \par\noindent\nopagebreak[4]
  \begin{problem}\label{prob:#2}\vspace{-0.6em}{\setlength{\fboxsep}{1pt}\colorbox{gray!17!white}{\textsc{#1}\index{problem!#1}}}\nopagebreak[4]\end{problem}\nopagebreak[4]\vspace{-0.6em}
  \par\noindent\hangindent=\parindent\textbf{#3}:  #4\nopagebreak[4]
  \par\noindent\hangindent=\parindent\textbf{#5}:  #6
  \par\bigskip
  \endgroup
  \end{samepage}
}

\DeclareMathOperator{\dist}{dist}

\newcommand{\I}{\mathcal{I}}
\newcommand{\yes}{\textnormal{\texttt{yes}}}
\newcommand{\no}{\textnormal{\texttt{no}}}
\newcommand{\RD}{$(\Rightarrow)\quad$}
\newcommand{\LD}{$(\Leftarrow)\quad$}
\newcommand{\tss}[1]{\textsuperscript{#1}}

\newcommandx{\decprobX}[5][2=Input,4=Question]{%
  \begingroup
  \par\medskip
  \noindent \colorbox{gray!17!white}{\textsc{#1}\index{problem!#1}}\nopagebreak[4]
  \par\noindent\hangindent=\parindent\textbf{#2}:  #3\nopagebreak[4]
  \par\noindent\hangindent=\parindent\textbf{#4}:  #5
  \par  \medskip
  \endgroup
}

\newcommand{\N}{\mathbb{N}}
\newcommand{\Nzero}{\mathbb{N}_0}

\renewcommand{\O}{\mathcal{O}}

\newcommand{\prob}[1]{\textnormal{\textsc{#1}}}

\newcommand{\gbp}{Green Bridges Placement}

\newcommand{\GBP}{\prob{GBP}}

\newcommand{\CongbpTsc}{\prob{Connected \gbp}}
\newcommand{\CongbpAcr}{\prob{Connect \GBP}}
\newcommandx{\RgbpTsc}[2][1=$d$]{\prob{#1-Reach \gbp}}
\newcommandx{\RgbpAcr}[2][1=$d$]{\prob{#1-Reach \GBP}}
\newcommandx{\CgbpTsc}[2][1=$d$]{\prob{#1-Closed \gbp}}
\newcommandx{\CgbpAcr}[2][1=$d$]{\prob{#1-Closed \GBP}}
\newcommandx{\DgbpTsc}[2][1=$d$]{\prob{#1-Diamater \gbp}}
\newcommandx{\DgbpAcr}[2][1=$d$]{\prob{#1-Diam \GBP}}

\newcommand{\cocl}[1]{\ensuremath{\operatorname{#1}}}
\newcommand{\W}[1]{\cocl{W[#1]}}

\newcommand{\NP}{\cocl{NP}}
\newcommand{\FPT}{\cocl{FPT}}

\newcommand{\coNP}{\cocl{coNP}}
\newcommand{\XP}{\cocl{XP}}
\newcommand{\poly}{\cocl{poly}}

\newcommand{\NPincoNPslashpoly}{\ensuremath{\NP\subseteq \coNP/\poly}}

\newcommand{\calI}{\mathcal{I}}

\newcommand{\calH}{\mathcal{H}}

\newcommand{\calF}{\mathcal{F}}

\newcommand{\tref}[1]{\scriptsize{(\Cref{#1})}}

\DeclareMathOperator{\diam}{diam}
\DeclareMathOperator{\indeg}{indeg}
\DeclareMathOperator{\outdeg}{outdeg}

\newcommand{\ceq}{\ensuremath{\coloneqq}}

\newcommand{\etal}{\textsl{et~al.}}

\newcommand{\cqed}{}
\newcommand{\lqed}{}
\newcommand{\squareforqed}{\qedhere}

\definecolor{lilla}{HTML}{750787}

\newcommand{\thecolor}{cyan!50!green}

\usepackage{makecell}

\usepackage{etoolbox}
\newcommand{\ExternalLink}{%
    \tikz[x=1.2ex, y=1.2ex, baseline=-0.05ex]{%
        \begin{scope}[x=1ex, y=1ex]
            \clip (-0.1,-0.1) 
                --++ (-0, 1.2) 
                --++ (0.6, 0) 
                --++ (0, -0.6) 
                --++ (0.6, 0) 
                --++ (0, -1);
            \path[draw, 
                line width = 0.5, 
                rounded corners=0.5] 
                (0,0) rectangle (1,1);
        \end{scope}
        \path[draw, line width = 0.5] (0.5, 0.5) 
            -- (1, 1);
        \path[draw, line width = 0.5] (0.6, 1) 
            -- (1, 1) -- (1, 0.6);
        }
    }

\makeatletter
  \def\abstractname{Abstract.}
  \renewenvironment{abstract}{%
      \if@twocolumn
        \section*{\abstractname}%
      \else
        \small
        \quotation
	\noindent{\bfseries\abstractname}%
      \fi}
      {\if@twocolumn\else\endquotation\fi}
\makeatother
\title{\Large \bf Placing Green Bridges Optimally, with a Multivariate~Analysis}
\author{Till Fluschnik\thanks{Supported by DFG, project TORE (NI/369-18).}~}
\author{Leon Kellerhals}
\affil{\small
	Technische Universit\"at Berlin, Faculty~IV, Institute of Software Engineering and Theoretical Computer Science, Algorithmics and Computational Complexity.
}

\date{}

\begin{document}

\maketitle

\begin{abstract}
We study the problem of placing wildlife crossings, 
such as green bridges,
over human-made obstacles to challenge habitat fragmentation.
The main task
herein is,
given
a graph describing habitats or routes of wildlife animals and possibilities of building green bridges,
to find a low-cost placement of green bridges that connects the habitats.
We develop three problem models for this task
and study them from a computational complexity 
and parameterized algorithmics perspective.
 
\noindent
\emph{Keywords.}
wildlife crossings $\cdot$
computational sustainability $\cdot$
connected subgraphs $\cdot$
NP-hardness $\cdot$
parameterized algorithmics

\end{abstract}

\section{Introduction}
\label{sec:intro}

Sustainability 
is
a major concern 
impacting today's
politics,
economy,
and industry.
Accordingly,
sustainability sciences are well-established by now.
Yet,
the \emph{interdisciplinary} scientific field ``computational sustainability''~\cite{Gomes09,Gomes09a},
which combines practical and theoretical computer science with sustainability sciences,
is quite young.
For instance,
the Institute for Computational Sustainability 
at Cornell University was founded in 2008,
the 1st International Conference on Computational Sustainability 
(CompSust'09) 
took place in 2009,
and
special tracks on
computational sustainability and AI
were established
at
AAAI 
\cite{AAAI10} 
and IJCAI
\cite{IJCAI13}.
This work
contributes to
computational sustainability:
We model problems of elaborately placing wildlife crossings
and give complexity-theoretical and algorithmic analysis for each.
Wildlife crossings are constructions (mostly bridges or tunnels~\cite{VanDerReeHMM09}) that allow wildlife animals to safely cross human-made transportation lines (e.g., roads). 
We will refer to wildlife crossings as~\emph{green bridges}.

There are numerous reports on wildlife-vehicle collisions~\cite{Shilling21,VanDerGriftSRG17,HuijserMHKCSA08}.
Huijser~\etal{} \cite{HuijserMHKCSA08} identify several endangered animal species suffering from high road mortality
and estimate the annual cost associated with wildlife-vehicle collisions with around 8~billion US dollars.
Wildlife fencing with wildlife crossings
can reduce collisions by over 80\%~\cite{HuijserMHKCSA08},
enables populations to sustain~\cite{SawayaKC14},
and are thereby among the most cost-effective~\cite{HuijserDCAM09}.
The implementation,
though,
is a delicate problem, as depicted by Huijser~\etal~\cite[p.\;\!16]{HuijserMHKCSA08}:

\begin{quote}
The location, type, and dimensions of wildlife
crossing structures must be carefully planned with regard to the species and surrounding
landscape. 
For example, grizzly bears, deer, and elk tend to use wildlife overpasses to a greater
extent than wildlife underpasses, while black bears and mountain lions use underpasses more
frequently than overpasses. 
In addition, different species use different habitats, influencing their
movements and where they want to cross the road.
\end{quote}

Apart from these delicacies, another challenge is to obtain good data about the specific areas inhabited by a species~\cite{ZhengLY19}:
While it is arguably easier to answer whether some animal species habitates a certain patch of land in the positive,
it seems more challenging to rule it out.
Clearly, high data quality is a crucial for deciding on where to place green bridges.

In this work,
we consider the task of (re-)connecting habitats under varying connectivity requirements
by placing as few green bridges as possible, thus minimizing the cost.
We assume to be given a set of land patches which are disconnected by roads, the set of inhabited patches for each animal, and possible locations for green bridges, each of which connects two patches.
This is canonical to model as a graph: vertices represent the land patches, edges represent the possible locations for green bridges, and for each animal species we are given a vertex subset of the inhabited patches.
The goal in the model now is to find an edge set that sufficiently connects the habitats of each species.

In particular,
we comparatively 
study in terms of computational complexity and parameterized algorithmics
the following three different (families of) decision problems.%
\footnote{The $d$-th power $G^d$ of a graph~$G$ contains edge~$\{v,w\}\in\binom{V(G)}{2}$ if and only if~$\dist_G(v,w)\leq d$.}
\decprobX{$\Pi$ \gbp{} ($\Pi$ \GBP{})}
{An undirected graph~$G=(V,E)$,
a set~$\calH=\{V_1,\dots,V_r\}$ of habitats where~$V_i\subseteq V$ for all~$i\in\set{r}$, 
and~$k\in\Nzero$.}
{Is there an edge set~$F\subseteq E$ with~$|F|\leq k$ such that for every~$i\in\set{r}$, 
it holds that~$V_i\subseteq V(G[F])$ and

\smallskip
\noindent
{
\setlength{\tabcolsep}{7pt}
\begin{tabular}{@{\hspace{1em}}rlll@{}}
 $\Pi\equiv{}$\prob{$d$-Reach}: & $G[F]^d[V_i]$ is connected? & (\cref{prob:rgbp}) & (\cref{sec:rgbp}) \\
 $\Pi\equiv{}$\prob{$d$-Closed}: & $G[F]^d[V_i]$ is a clique? & (\cref{prob:cgbp}) &  (\cref{sec:cgbp})\\
 $\Pi\equiv{}$\prob{$d$-Diam(eter)}: & $\diam(G[F][V_i])\leq d$? & (\cref{prob:dgbp}) & (\cref{sec:dgbp})
\end{tabular}
}}

Our problems address both the challenge in obtaining high quality data as well as the question to what connectivity is sufficient.
Connectivity is addressed by the different requirements on the solution:
While \RgbpAcr{} simply ensures connectivity of each habitat along length-$d$ paths,
\CgbpAcr{} additionally requires every two patches of each habitat to be connected by such a path.
The latter is also true for \DgbpAcr{}, 
which additionally requires that such a path only uses the habitat's patches. 
In this sense,
\DgbpAcr{} generalizes \RgbpAcr[1]{}.
Moreover, 
\DgbpAcr{} and \CgbpAcr{} are equivalent for~$d=1$.
See~\cref{fig:relationship}
for relationships between the problems.

As for the data quality,
recall that it is arguably easier to tell with sufficient certainty that some animal species inhabits a certain area, but harder to rule it out with the same certainty, especially for areas that are adjacent to habitated areas.
This property is captured very well by \RgbpAcr{} and \CgbpAcr{}.
Herein, one should choose $d$ antiproportionally to the data quality.
For instance, with perfect data quality, that is, perfect knowledge about each species' habitat, one may choose~$d=1$ (and hence, \DgbpAcr{} is also amenable).
Imperfect data quality is reflected by a choice of~$d > 1$.
Here, we relax the connectivity constraints and allow for ``hops'' within the connected habitat.
If for example $d=2$ and a possibly uninhabited area $v$ is adjacent to two inhabited areas $u$ and~$w$, then $u$ and~$w$ may be connected by~$\{u, v\}$ and~$\{v, w\}$, thus ``hopping'' over~$v$.

\begin{figure}[t]\centering
\begin{tikzpicture}[xscale=1.2,yscale=1.7]
    \def\xr{1}
    \def\yr{1}
    \node (con) at (0,-1.40*\yr)[]{\hyperref[prob:congbp]{\CongbpAcr}};
    \node (r) at (-3*\xr,-1*\yr)[]{\hyperref[prob:rgbp]{\prob{Reach \GBP{}}}};
    \node (c) at (3*\xr,-1*\yr)[]{\hyperref[prob:cgbp]{\prob{Closed \GBP{}}}};
    \node (r1) at (-4*\xr,-1.75*\yr)[]{\RgbpAcr[1]{}};
    \node (c1) at (4*\xr,-2*\yr)[]{\CgbpAcr[1]{}};
    \node (d) at (-1.25*\xr,-1.75*\yr)[]{\hyperref[prob:dgbp]{\prob{Diam \GBP{}}}};
    \node (d1) at (1.25*\xr,-2*\yr)[]{\DgbpAcr[1]{}};

    \draw[semithick,->] (con) to (r);
    \draw[semithick,->] (con) to (c);
    \draw[semithick,->] (con) to (r);
    \draw[semithick,->] (con) to (c);
    \draw[semithick,->] (r1) to (r);
    \draw[semithick,->] (c1) to (c);
    \draw[semithick,<->] (r1) to (d);
    \draw[semithick,->] (d1) to (d);
    \draw[semithick,<->] (d1) to (c1);
  \end{tikzpicture} 
  \caption{
      A diagram of interconnections between the problems (for the definition of \CongbpAcr{} see~\cref{prob:congbp}).
      An edge from problem~$A$ to problem~$B$ means that any solution to~$A$ is also a solution to~$B$.
      Problems with~$d$ omitted from the problem name require that there is a solution for some value of $d$.
  }
  \label{fig:relationship}
\end{figure}

\paragraph{Our Contributions.}
Our results are summarized in \cref{tab:results}.
\begin{table}[t]
 \caption{Overview of our results. 
 \NP-c., P, K, \W{1}-h., and~p-\NP-h. 
 stand for 
 \NP-complete,
 ``polynomial-size'',
 ``problem kernel'',
 \W{1}-hard,
 and para-\NP-hard,
 respectively.\newline
 \tss{a}(even on planar graphs or if~$\Delta=4$)
 \tss{b}(even on bipartite graphs with~$\Delta=4$ or graphs of diameter four)
 \tss{c}(even if~$r=1$ or if~$r=2$ and~$\Delta=4$)
 \tss{d}(even on bipartite graphs of diameter three and~$r=1$, 
 \emph{but} linear-time solvable when~$r+\Delta$ is constant)
 \tss{e}(admits a linear-size problem kernel if~$\Delta$ is constant)
 \tss{f}(linear-time solvable when~$r+\Delta$ is constant) 
 \tss{g}(even if~$r=1$)
 \tss{$\dagger$}(no polynomial problem kernel unless~\NPincoNPslashpoly)
 \tss{*}(but an~$\O(k^3)$-vertex problem kernel on planar graphs)
 \tss{$\ddagger$}(if~$r\geq 7$, linear-time solvable if~$r\leq 2$)
 }
 \label{tab:results}
 \centering
 \setlength{\tabcolsep}{3.25pt}
 \begin{tabular}{@{}p{0.12\textwidth}l|p{0.11\textwidth}|p{0.21\textwidth}p{0.13\textwidth}p{0.17\textwidth}|p{0.1\textwidth}@{}}\toprule
  Problem &  & Comput. & \multicolumn{3}{p{0.46\textwidth}|}{%
  Parameterized Algorithmics} & Ref.\\
  ($\Pi$~\GBP) &  & Complex. & $k$ & $r$ & $k+r$ &
  \\ \midrule\midrule
  \multirow{3}{*}{\makecell{\prob{$d$-Reach} \\ \tref{sec:rgbp}}} 
                              & $d=1$ & \NP-c.\tss{a} & $2k$-vertex K\tss{$\dagger$} & p-\NP-h.$^\ddagger$ & $O(rk+k^2)$ PK & \tref{ssec:1rgbp}\\
                              & $d=2$ & \NP-c.\tss{b}  & $O(k^k)$-vertex K\tss{$\dagger$,*} & p-\NP-h.\tss{c} & FPT\tss{$\dagger$} & \tref{ssec:2rgbp} \\
                              & $d\geq 3$ & \NP-c. & \XP, \W{1}-h. & p-\NP-h.\tss{c} & \XP, \W{1}-h. & \tref{ssec:3rgbp}\\
  \midrule
  \multirow{3}{*}{\makecell{\prob{$d$-Closed} \\ \tref{sec:cgbp}}} 
                              & $d=1$ & Lin.~time & --- & --- & --- & \tref{sec:cgbp} \\
                              & $d=2$ & \NP-c.\tss{d} & $O(k^k)$-vertex K\tss{$\dagger$,*} & p-\NP-h.\tss{e,g}  & FPT\tss{$\dagger$} & \tref{ssec:2cgbp}\\
                              & $d\geq 3$ & \NP-c. & \XP, \W{1}-h. & p-\NP-h.\tss{e,g} & \XP, \W{1}-h. & \tref{ssec:3cgbp}\\
  \midrule
  \multirow{2}{*}{\makecell{\prob{$d$-Diam} \\ \tref{sec:dgbp}}} & $d=1$ & Lin.~time & --- & --- & --- & \tref{sec:dgbp}\\
                              & $d=2$     & \NP-c.\tss{f} & $2k$-vertex K\tss{$\dagger$} & p-\NP-h.\tss{g} & $O(rk+k^2)$ PK & \tref{sec:dgbp} %
  \\\arrayrulecolor{black} \bottomrule
 \end{tabular}
\end{table}
We settle the classic complexity and parameterized complexity (regarding the number~$k$ of green bridges and the number~$r$ of habitats)
of the three problems.
While~\RgbpAcr{} is 
(surprisingly) 
already \NP-hard for~$d=1$ on planar or maximum degree~$\Delta=4$ graphs,
\CgbpAcr{} and \DgbpAcr{} become~\NP-hard for~$d\geq 2$,
but admit an $(r+\Delta)^{\O(1)}$-sized problem kernel 
and thus are linear time solvable if~$r+\Delta$ is constant.
All variants are para-\NP-hard when parameterized by $r$.
\RgbpAcr{} and \CgbpAcr{}
are  fixed-parameter tractable regarding~$k$ when~$d\leq 2$,
but become~\W{1}-hard (yet~\XP) regarding~$k$ and~$k+r$ when~$d>2$.
Additionally,
we prove that
\RgbpAcr{} admits an $rd$-approximation in~$\O(mn+rnd)$ time.

\paragraph{Further related work.}
Our problems deal with
finding (small) spanning connected subgraphs
obeying some (connectivity) constraints.
These problems are applicable in a wide range of areas and typically take the form of a special case or variant of \RgbpAcr[1]{}.
Areas include
computer networks~\cite{chockler2007pubsub},
social networks~\cite{angluin2015network},
graph drawing~\cite{brandes2012supports},
combinatorial auctions~\cite{conitzer2004auctions},
reconfigurable computing~\cite{fan2008interconnection},
vacuum technology~\cite{du1995interconnection},
and structural biology~\cite{agarwal2013inference}.

\RgbpAcr[1]{} on cliques is also known as the \prob{Subset Interconnection Design} problem: Given sets $V_1, \dots, V_r$, find a graph $G$ with $V(G) = V_1 \cup \dots \cup V_r$ with the minimum number of edges such that $G[V_i]$ is connected for each $i$.
This problem was first introduced by \cite{du1986optimization} and proven to be NP-hard by \cite{du1988interconnection}.
It was also studied in terms of its approximability~\cite{angluin2015network} and its parameterized complexity~\cite{chen2015interconnection}.

Closely related to our problems are also
Steiner multigraph problems~\cite{RicheyP86,Gassner10},
which were also studied in the context of wildlife corridor construction~\cite{LaiGSMCM11,lebras2013robust}.
Requiring small diameter 
appears also in the context of spanning trees~\cite{RaviSMRR96}
and Steiner forests~\cite{DingQ20}.
An edge-weighted version of
\DgbpAcr[4]{}
is proven to be \NP-hard even if there are only two different weights~\cite{Plesnik81}.
Kim \etal{} \cite{KimMMP20} study the problem of deleting few edges to augment a graph's diameter to a constant.
Gionis \etal{} \cite{gionis2017community} studied a variant of \DgbpAcr[2]{} in which for any solution $F$ and habitat $V_i$, $G[F][V_i]$ must induce a star, and gave an efficient approximation algorithm for it.
Herrendorf \cite{herrendorf2022sparsification} studied the same variant as well as the \RgbpAcr[1]{} problem (under a different name) in terms of their parameterized complexity.
 
\paragraph{Connecting habitats arbitrarily.}
The following obvious model \emph{just} requires that each habitat is connected.

\decprob{\CongbpTsc{} (\CongbpAcr{})}{congbp}
{An undirected graph~$G=(V,E)$,
a set~$\calH=\{V_1,\dots,V_r\}$ of habitats 
where~$V_i\subseteq V$ for all~$i\in\set{r}$, 
and an integer~$k\in\Nzero$.}
{Is there a subset~$F\subseteq E$ with~$|F|\leq k$ such that
for every~$i\in\set{r}$
it holds that in~$G[F]$ there exists a connected component containing~$V_i$?
}

\noindent
\CongbpAcr{}
with edge costs
is also known as \prob{Steiner Forest}~\cite{Gassner10}
and generalizes the well-known \NP-hard \prob{Steiner Tree} problem. 
Gassner~\cite{Gassner10} proved \prob{Steiner Forest} to be \NP-hard 
even if every so-called terminal net contains two vertices,
if the graph is planar and has treewidth three,
and if there are two different edge costs,
each being upper-bounded linearly in the instance size.
It follows that \CongbpAcr{} is also \NP-hard in this case.
Bateni~\etal~\cite{BateniHM11} proved that \prob{Steiner Forest} is polynomial-time solvable on graphs of treewidth two
and admits approximation schemes on planar and bounded-treewidth graphs.

From a modeling perspective,
solutions for \CongbpAcr{} may be highly scattered:
Patches of the same species' habitat may be arbitrarily far away from another;
thus, to reach another patch of their habitat, animals may need to take long walks through areas of their habitats when only using green bridges to cross streets.
It is likely that species with scattered habitats will not make use of the green bridges.
With our models we avoid such solutions.

\section{Preliminaries}
\label{sec:prelims}

Let~$\N$ and~$\Nzero$ be the natural numbers without and with zero,
respectively.
We use basic definitions from graph theory~\cite{Diestel} and
parameterized algorithmics~\cite{cygan2015parameterized}.

\paragraph{Graph Theory.}

Let~$G=(V,E)$ be an undirected graph with vertex set~$V$ and edge set~$E\subseteq \binom{V}{2}$.
We also denote by~$V(G)$ and~$E(G)$ the vertices and edges of~$G$,
respectively.
For~$V'\subseteq V$,
let~$G[V']=(V',E\cap \binom{V'}{2})$ denote the graph~$G$ induced by a vertex set~$V'$. 
For~$F\subseteq E$ let~$V(F)\ceq \{v\in V\mid \exists e\in F :\: v\in e\}$ and
$G[F]\ceq (V(F),F)$ denote the graph~$G$ induced by the edge set~$F$.
A path~$P$ is a graph with~$V(P) \ceq \{v_1, \ldots, v_n\}$ and~$E(P) \ceq \{\{v_i,v_{i+1}\} \mid 1 \le i < n \}$.
The length of the path~$P$ is $|E(P)|$.
The distance~$\dist_G(v,w)$ between vertices~$v,w \in V(G)$ is the length of the shortest path between~$v$ and~$w$ in~$G$.
The diameter~$\diam(G)$ is the length of longest shortest path over all vertex pairs.
For~$p\in \N$,
the graph~$G^p$ is the $p$-th power of~$G$ 
containing the vertex set~$V$ and edge set~$\{\{v,w\}\in \binom{V}{2}\mid \dist_G(v,w)\leq p\}$.
For~$F\subseteq E$,
$V'\subseteq V$,
and~$d\in\N$,
the graph~$G[F]^d[V']$ is understood as~$((G[F])^d)[V']$.
Let~$N_G(v)\ceq \{w\in V\mid \{v,w\}\in E\}$ be the (open) neighborhood of~$v$,
and $N_G[v]\ceq N_G(v)\cup\{v\}$ be the closed neighborhood of~$v$.
For~$p\in\N$,
let~$N_G^p(v)\ceq \{w\in V\mid \{v,w\}\in E(G^p)\}$ be the (open) $p$-neighborhood of~$v$,
and $N_G^p[v]\ceq N_G^p(v)\cup\{v\}$ be the closed $p$-neighborhood of~$v$.
Two vertices~$v,w\in V$ are called twins if~$N_G(v)=N_G(w)$.
The (vertex) degree~$\deg_G(v)\ceq |N_G(v)|$ of~$v$ is the number of its neighbors.
The maximum degree~$\Delta(G)\ceq \max_{v\in V}\deg_G(v)$ is the maximum over all (vertex) degrees.

\section{Connecting Habitats with a Patch at Short Reach}\looseness=1
\label{sec:rgbp}

The following problem ensures that any habitat patch can reach the other patches via patches of the same habitat and short strolls over ``foreign'' ground.

\decprob{\RgbpTsc{} (\RgbpAcr{})}{rgbp}
{An undirected graph~$G=(V,E)$,
a set~$\calH=\{V_1,\dots,V_r\}$ of habitats where~$V_i\subseteq V$ for all~$i\in\set{r}$, 
and an integer~$k\in\Nzero$.}
{Is there a subset~$F\subseteq E$ with~$|F|\leq k$ such that
for every~$i\in\set{r}$
it holds that~$V_i\subseteq V(G[F])$ and~$G[F]^d[V_i]$ is connected?
}

\begin{theorem}
\label{thm:rgbp}
 \RgbpTsc{} is
 \begin{compactenum}[(i)]
  \item if~$d=1$, \NP-hard even on planar graphs, 
    or if~$r\geq 7$ but solvable in linear time if~$r \le 2$;
  \item if~$d=2$, \NP-hard even on graphs with maximum degree four and~$r=2$ or graphs with diameter four and~$r=1$, and in~\FPT{} regarding~$k$;
  \item if~$d\geq 3$, \NP-hard and~\W{1}-hard regarding~$k+r$.
 \end{compactenum}
 Moreover, \RgbpAcr{} admits an~$rd$-approximation of the minimum number of green bridges in~$\O(mn+rnd)$ time.
\end{theorem}

We will first present the approximation algorithm. 
Afterwards, we will present the results in (i)-(iii) in the order above.

\subsection{An \texorpdfstring{$(r\cdot d)$}{(r·d)}-Approximation for~\texorpdfstring{\RgbpAcr{}}{d-Reach GBP}}

\noindent
In this section we will present the approximation algorithm of \cref{thm:rgbp}.
The approximation algorithm computes for every habitat~$V_i$ a spanning tree in~$G^d[V_i]$,
and adds the edges of the corresponding paths to the solution~$F$.
Each of the spanning trees then is a~$d$-approximation for just the one habitat,
hence the union of the spanning trees is an~$rd$-approximation for all habitats.

\begin{lemma}
 \label{lem:d-approx}
 For~$r=1$,
 \RgbpAcr{} admits a~$d$-approximation 
 of the minimum number of green bridges
 in~$\O(mn)$ time.
\end{lemma}
\begin{proof}
	We start off by computing in~$\O(mn)$ time the graph~$H \ceq G^d$ as well as for every edge~$e = \{u, v\} \in E(H)$ the corresponding path~$P_e$ from~$u$ to~$v$ of length at most~$d$ in~$G$.
	If~$H[V_1]$ is not connected, then return~\no{}.
	If not,
	then compute a minimum spanning tree~$T \subseteq H[V_1]$ in~$\O(n \log n)$ time.
	For each edge~$e = \{u, v\} \in E(T)$ compute in~$\O(m)$ time the corresponding path~$P_e \subseteq G$ from~$u$ to~$v$ of length at most~$d$.
	Finally, return the set~$F \ceq \bigcup_{e \in E(T)} E(P_e)$, computable in~$\O(m)$ time.
	Clearly, $G[F]^d[V_1]$ is connected.
	As a minimum solution~$F^*$ has at least~$|V_1|-1$ edges,
	and every path~$P_e$ consists of at most~$d$ edges,
	\begin{linenomath*}
	\[
		|F| = |\bigcup_{e\in E(T)} E(P_e)| \le \sum_{e \in E(T)} E(P_e) \le (|V_1|-1)\cdot d \le d|F^*|. \qedhere
	\]
	\end{linenomath*}
\end{proof}
\begin{proposition}
	\label{prop:rgbp-approx}
	\RgbpAcr{} admits an~$rd$-approximation 
	of the minimum number of green bridges 
	in $\O(mn + rnd)$ time.
\end{proposition}
\begin{proof}
	We initially compute the shortest paths between all vertex pairs in~$G$ in~$O(mn)$ time.
	We obtain the graph~$H \ceq G^d$ as a byproduct.
	If for some~$i \in \set{r}$, $H[V_i]$ is not connected,
	then return~\no{}.
	If not, then 
	compute for each~$i \in \set{r}$ a spanning tree~$T_i$ of~$H[V_i]$, or return~\no{} if~$H[V_i]$ is not connected.
	Let~$F_i \subseteq E(G)$ be the edge set corresponding to~$T_i$ as in the proof of \cref{lem:d-approx}.
	As~$G[F_i]^d[V_i]$ is connected,
	$F \ceq \bigcup_{i=1}^r F_i$ is a solution.

	Note that each of the~$r$ spanning trees~$T_i$ contains at most~$n$ edges,
	and for each of these edges~$e \in F_i$ we can determine the corresponding paths~$P_e \subseteq G$
	of length at most~$d$ in~$\O(d)$ time.
	We obtain an overall running time of~$\O(mn + rnd)$.

	As for the approximation ratio, 
	let~$F^*$ be a minimum solution, 
	and for every~$i \in \set{r}$ let~$F^*_i \subseteq E(G)$ be a minimum-size edge set such that~$G[F^*_i]^d[V_i]$ is connected.
	As~$|F^*| \ge \max_{i\in\set{r}} |F^*_i|$,
	we have
	\begin{linenomath*}
	\[
		|F| \le \sum_{i=1}^r |F_i| \le \sum_{i=1}^r d|F^*_i| \le r \cdot d|F^*|. \qedhere
	\]
	\end{linenomath*}
\end{proof}

\subsection{When a next habitat is directly reachable (\texorpdfstring{$d=1$}{d=1})}
\label{ssec:1rgbp}

Recall that setting~$d=1$ may reflect perfect knowledge about the habitats.
In this case,
we want that in~$G[F]$, 
each habitat~$V_i$ forms a connected component.

Du and Miller \cite{du1988interconnection} showed that \RgbpAcr[1]{} is NP-hard even when the input graph is complete.
We give two reductions that show NP-hardness in some restricted cases.
From the second reduction we can also derive that presumably there is no polynomial kernel with respect to the budget~$k$.
Lastly, we show that if there are only two habitats, then the problem can be solved in linear time.

We start with proving that \RgbpAcr[1]{} is \NP-hard on series-parallel graphs.
As every series-parallel graph is planar, we also obtain the same hardness result for planar graphs.
Further, the provided reduction also shows that the problem is unlikely to admit a kernel whose size is bounded polynomially in the parameter.

\begin{proposition}
	\label{prop:1rgbp-planar}
	\RgbpAcr[1]{} is \NP-hard and, unless $\NPincoNPslashpoly$, admits no problem kernel of size~$k^{\O(1)}$, even on series-parallel graphs.
\end{proposition}

We will give a linear parametric transformation from the following problem:

\decprob{Hitting Set (HS)}{hittingset}
{A universe~$U$, a set~$\calF\subseteq 2^U$ of subsets of~$U$, and an integer~$k$.}
{Is there a \emph{hitting set}~$U' \subseteq U$ with $|U'|\leq k$ such that for all~$F\in\calF$ we have~$F\cap U' \ne \emptyset$?}

Note that \prob{Hitting Set} admits no problem kernel of size polynomial in~$|U|$ unless~$\NPincoNPslashpoly$~\cite{DomLS14}.

\begin{construction}
 \label{constr:1rgbp-planar}
 For an instance~$\I=(U,\calF,k)$ of~\prob{Hitting Set}
 with $U=\set{n}$ and $\calF=\{F_1, \dots, F_m\}$,
 we construct an instance~$\I'\ceq (G',\calH,k')$ with $k'\ceq n+k$
 and habitat set $\calH=\{S,V_1,\dots,V_m\}$
 as follows
 (see~\cref{fig:1rgbp-planar} for an illustration).
\begin{figure}[t]
  \centering
    \begin{tikzpicture}
     \contourlength{0.09em}

      \def\xr{1}
      \def\yr{1}
      \def\xs{1.5}
      \def\ys{1}
      \def\dout{25}
      \def\din{155}
      \tikzstyle{xnode}=[circle,fill,scale=0.6,draw,color=\thecolor]
      \tikzstyle{xnodex}=[label={[xshift=-0.5*\xr cm]0:$\cdots$},inner sep=5pt,color=\thecolor]
      \tikzstyle{xnodef}=[circle,scale=0.6,draw,color=\thecolor,fill=white]
      \tikzstyle{xedge}=[thick,-,color=\thecolor]

      \foreach \x/\y in {1/xnode,2/xnodex,3/xnode,4/xnodex,5/xnode,6/xnodex,7/xnode}{
          \node (a\x) at (\x*\xs*\xr,0)[\y]{};
      }
      \foreach \x/\y in {1/$x_1$,3/$x_i$}{
          \node at (a\x)[anchor=east]{\y};
      }
      \foreach \x/\y in {5/$x_j$,7/$x_n$}{
          \node at (a\x)[anchor=west]{\y};
      }
      
      \node (a) at (4*\xs*\xr,1*\yr)[xnode]{};
      \node (b) at (4*\xs*\xr,-1*\yr)[xnode]{};
      \foreach \x/\y in {1,...,7}{
        \draw[xedge,color=red!50!black] (a\x) to (a);
        \draw[xedge] (a\x) to (b);
      }
      
      \newcommand{\ltb}[3]{
        \node at (#1)[label={[align=center,
            font=\scriptsize,color=black]\contour*{white}{#2}},
            label={[align=center,font=\scriptsize,color=black]-90:\contour*{white}{#3}}]{};
      }
      \ltb{a1}{$\in S,V_p$~~~}{};
      \ltb{a3}{$\in S,V_p,V_q$~~~}{};
      \ltb{a5}{~~~$\in S,V_p,V_q$}{};
      \ltb{a7}{~~~$\in S,V_q$}{};
      
      \ltb{a}{$\in S,V_1,\dots,V_m$}{$s$};
      \ltb{b}{$t$}{$\in V_1,\dots,V_m$};
    
    \end{tikzpicture}
  \caption{Illustration to~\cref{constr:1rgbp-planar} for \RgbpAcr[1]{} on series-parallel (and thus planar) graphs. 
  In this example,
  there are e.g.~$F_p\supseteq\{1,i,j\}$ and~$F_q\supseteq\{i,j,n\}$.
  In case of a \yes-instance,
  the red-colored edges are in every solution~(\cref{obs:1rgbp-planar}).
  }
  \label{fig:1rgbp-planar}
 \end{figure}

 Add to~$G'$ the vertex set
 $V_U \ceq \{x_i \mid i \in U\}$
 as well as the two vertices~$s$ and~$t$,
 and the edge sets
 $E^* \ceq \bigcup_{i=1}^n\{\{s, x_i\}\}$
 and~$E_U \ceq \bigcup_{i=1}^n\{\{x_i, t\}\}$.
 Finally,
 let~$S\ceq\{s\}\cup\bigcup_{i=1}^n \{x_i\}$,
 and for each~$F_j \in \calF$
 let~$V_j \ceq \{s, t\}\cup\bigcup_{i \in F_j} \{x_i\}$.
 \cqed
\end{construction}

\begin{observation}
	The graph~$G'$ constructed in \cref{constr:1rgbp-planar} is planar and series-parallel.
\end{observation}

\begin{observation}
 \label{obs:1rgbp-planar}
 Let~$\I'$ be a \yes-instance.
 Then
 every solution~$F$ contains all edges in~$E^*$.
\end{observation}

{
\begin{proof}
	By construction, $G[S]$ is a star with center~$s$.
	Hence, all edges in~$G[S]$ are contained in every solution.
	Since~$E^* = E(G[S])$, the claim follows.
	\lqed
\end{proof}
}

\begin{lemma}
 \label{lem:1rgbp-planar}
 Let~$\I'$ be the instance obtained from applying \cref{constr:1rgbp-planar} to an instance~$\I$.
 Then,
 $\I$ is a \yes-instance if and only if~$\I'$ is a \yes-instance.
\end{lemma}

\begin{proof}
  \RD{}
  Let~$U' \subseteq U$ be a solution for instance~$\I$.
  We claim that~$F\ceq E^* \cup \bigcup_{i \in U'}\{\{x_i,t\}\}$
  is a solution for~$\I'$.
  Note that~$|F| \le n+k$.
  Observe that~$G'[F][S]$ is connected.
  Suppose now that there is~$V_j$ such that~$G'[F][V_j]$
  is not connected.
  Let~$F_j$ be the corresponding set.
  Since~$E^*\subseteq F$,
  none of the edges~$\{\{x_i, t\} \mid i\in F_j\}$
  are contained in~$F$.
  It follows that~$F_j \cap U'=\emptyset$, 
  contradicting the fact that~$U'$ is a solution for~$\I$.
  
  \LD{}
  Let~$F$ be a solution to~$\I'$.
  By \cref{obs:2rgbpplanar} we know that~$E^*\subseteq F$.
  We claim that~$U' \ceq \{i\in U \mid \{x_i, t\} \in F\}$
  is a solution for~$\I$.
  Clearly~$|U'| \le k$.
  Suppose~$U'$ is not a solution.
  Then there is an~$F_j \in \calF$ with~$F_j \cap U'=\emptyset$.
  But then~$G'[F][V_j]$ is not connected,
  a contradiction.
	\lqed
\end{proof}

Next, we prove that \RgbpAcr[1]{} is \NP-hard even if we are given a constant number of habitats.

\begin{proposition}
 \label{prop:1rgbpNPhconsthab}
 \RgbpAcr[1]{} is \NP-complete even if~$r=7$.
\end{proposition}

We reduce from the following \NP-hard problem.

\decprob{Directed Hamiltonian Path (DHP)}{dhp}
{A directed graph~$D=(W,A)$ and two distinct vertices~$s,t\in W$ such that $\outdeg(t)=\indeg(s)=0$.}
{Is there an~$s$-$t$ path that visits every vertex exactly once?}

We first recall a well-known reduction to \prob{Hamiltonian Path (HP)},
the undirected variant. 
Then, we reduce HP to \RgbpAcr[1]{}.
For both constructions,
we refer to \cref{fig:1rgbp-paranp} for an illustrative example.

\begin{construction}
 \label{constr:1rgbpNPhconsthab-A}
 Construct the undirected graph~$G'=(V',E')$ as follows.
 For each vertex~$v\in W\setminus\{s,t\}$,
 $G'$ contains the path~$P_v=(v_{\rm in},v,v_{\rm out})$.
 Moreover, it contains the paths~$P_s=(s,s_{\rm out})$ and~$P_t=(t_{in},t)$.
 For every arc~$(v,w)\in A$,
 add the edge~$(v_{\rm out}, w_{\rm in})$.
 \cqed
\end{construction}

Recall the following.

\begin{observation}
 $(D,s,t)$ is a \yes-instance of \prob{DHP}
 if and only if
 $(G',s,t)$,
 obtained from~$(D,s,t)$ using~\cref{constr:1rgbpNPhconsthab-A},
 is a \yes-instance of \prob{HP}.
\end{observation}

Next,
we construct the instance of \RgbpAcr[1]{} from~$G'$.

\begin{construction}
	\label{constr:1rgbpNPhconsthab-B}
	Let~$G'=(W',E')$ be the graph obtained from~$(D=(W,A),s,t)$ using~\cref{constr:1rgbpNPhconsthab-A}.
	We now construct the graph~$G=(V,E)$ from~$G'$ with habitat set 
	\[ \calH=\{X_{\rm out},X_{\rm in},X_{\rm out}^*,X_{\rm in}^*,V_{\rm all},Y_{\rm out},Y_{\rm in}\} \] as follows.
	Add the new vertices $x_{\rm out},x_{\rm in},y_{\rm out},y_{\rm in}$,
	that is,
	let \[V\ceq V'\cup \{x_{\rm out},x_{\rm in},y_{\rm out},y_{\rm in}\} .\]
	Moreover,
	make~$x_{\rm out}$ adjacent to~$v_{\rm out}$ for each~$v\in W\setminus\{t\}$,
	make~$y_{\rm out}$ adjacent to~$v_{\rm out}$ for each~$v\in W\setminus\{t\}$,
	make~$x_{\rm in}$ adjacent to~$v_{\rm in}$ for each~$v\in W\setminus\{s\}$,
	make~$y_{\rm in}$ adjacent to~$v_{\rm in}$ for each~$v\in W\setminus\{s\}$,
	Next,
	let~$V_{\rm all}\ceq W'$ and
	\begin{linenomath*}
	\begin{align*}
		X_{\rm out}^* &\ceq \{x_{\rm out}\} \cup \bigcup_{v\in W\setminus\{t\}}\{v_{\rm out}\}, 
		& X_{\rm out} &\ceq X_{\rm out}^* \cup \bigcup_{v\in W\setminus\{s\}}\{v_{\rm in}\},  \\
		X_{\rm in}^* &\ceq \{x_{\rm in}\}  \cup \bigcup_{v\in W\setminus\{s\}}\{v_{\rm in}\}, 
		& X_{\rm in} &\ceq  X_{\rm in}^* \cup \bigcup_{v\in W\setminus\{t\}}\{v_{\rm out}\}, \\
		Y_{\rm out} &\ceq \{y_{\rm out}\} \cup \bigcup_{v \in W \setminus\{t\}} \{v_{\rm out}, v\},\text{ and} 
		& Y_{\rm in} &\ceq \{y_{\rm in}\} \cup \bigcup_{v \in W \setminus\{s\}} \{v_{\rm in}, v\}.
	\end{align*}
	\end{linenomath*}
	Finally,
	let~$k\ceq 2(n-2)+2 + 4(n-1) + (n-1) = 7(n-1)$, where~$n = |W|$.
	\cqed
\end{construction}

\begin{figure}[t]
  \centering
    \begin{tikzpicture}
      \contourlength{0.09em}

      \def\xr{0.95}
      \def\yr{1}
      \def\xs{1.15}
      \def\ys{1}
      \def\dout{25}
      \def\din{155}
      \tikzstyle{xnode}=[circle,fill,scale=0.6,draw,color=\thecolor]
      \tikzstyle{xnodex}=[scale=0.8,draw,color=\thecolor,fill=yellow]
      \tikzstyle{xnodey}=[scale=0.8,draw,color=\thecolor,fill=magenta]
      \tikzstyle{xnodef}=[circle,scale=0.6,draw,color=\thecolor,fill=white]
      \tikzstyle{xedge}=[-,color=\thecolor]
      \tikzstyle{xedgex}=[ultra thick,-,color=red!50!black]
      \tikzstyle{xarc}=[thick,->,>=latex,color=\thecolor]
      
      \newcommand{\gprime}[1]{
       \node (s) at (0*\xr,3*\yr)[xnode]{};
       \node (s1out) at (0*\xr,2.5*\yr)[xnode]{};
       \node (a1in) at (0*\xr,1.5*\yr)[xnode]{};
       \node (a) at (0*\xr,1*\yr)[xnode]{};
       \node (a1out) at (0*\xr,0.5*\yr)[xnode]{};
       \node (b1in) at (0*\xr,-0.5*\yr)[xnode]{};
       \node (b) at (0*\xr,-1*\yr)[xnode]{};
       \node (b1out) at (0*\xr,-1.5*\yr)[xnode]{};
       \node (t1in) at (0*\xr,-2.5*\yr)[xnode]{};
       \node (t) at (0*\xr,-3*\yr)[xnode]{};
       
       \foreach \x/\y in {s/a,a/b,a/t,b/t}{\draw[xedge] (\x1out) to (\y1in);}
       \draw[xedge] (b1out) to [out=45,in=-45](a1in);
       \draw[xedge] (a1out) to [out=-45,in=45](t1in);
       \foreach \x in {a,b,t}{\draw[xedge] (\x) to (\x1in);}
       \foreach \x in {s,a,b}{\draw[xedge] (\x) to (\x1out);}
       \foreach \x/\y/\z in {s/$s$/west,s1out/$s_{\rm out}$/west,a1in/$a_{\rm in}$/west,a/$a$/west,a1out/$a_{\rm out}$/west,
       b1in/$b_{\rm in}$/west,b/$b$/west,b1out/$b_{\rm out}$/west,
       t1in/$t_{\rm in}$/west,t/$t$/west}
       {\node at (\x)[inner sep=5pt,anchor=\z]{\contour*{white}{\y}};}
       
       \ifnum#1=1
         \node (x1out) at (-1*\xr,1*\yr)[xnode]{};
         \node (y1out) at (-1*\xr,0.5*\yr)[xnode]{};
         \node (x1in) at (-1*\xr,-0.5*\yr)[xnode]{};
         \node (y1in) at (-1*\xr,-1*\yr)[xnode]{};
         \foreach \x/\y/\z in {x1out/$x_{\rm out}$/east,y1out/$y_{\rm out}$/east,x1in/$x_{\rm in}$/east,y1in/$y_{\rm in}$/east}
	 {\node at (\x)[inner sep=5pt,anchor=\z]{\contour*{white}{\y}};}
         \foreach \x in {x,y}{\foreach \y in {s,a,b}{\draw[xedge] (\x1out) to (\y1out);}}
         \foreach \x in {x,y}{\foreach \y in {a,b,t}{\draw[xedge] (\x1in) to (\y1in);}}
       \fi
      }
      
      \newcommandx{\xlabel}[2][1=-1.75]{\node at (#1*\xr,3.25*\yr)[]{#2};}
      
      \begin{scope}
       \xlabel[-0.5]{(a)}
       \node (s) at (0*\xr,3*\yr)[xnode]{};
       \node (a) at (0*\xr,1*\yr)[xnode]{};
       \node (b) at (0*\xr,-1*\yr)[xnode]{};
       \node (t) at (0*\xr,-3*\yr)[xnode]{};
       \foreach \x/\y/\z in {s/a,b/t}{\draw[xarc] (\x) to (\y);}
       \draw[xarc] (a) to [bend left](b);
       \draw[xarc] (b) to [bend left](a);
       \draw[xarc] (a) to [out=-45,in=45](t);
       \foreach \x/\y/\z in {s/$s$/west,a/$a$/west,b/$b$/west,t/$t$/west}
       {\node at (\x)[inner sep=5pt,anchor=\z]{\contour*{white}{\y}};}
      \end{scope}
      
      \begin{scope}[xshift=2*\xs cm]
       \xlabel[-0.5]{(b)}
       \gprime{0}
      \end{scope}
      
      \begin{scope}[xshift=5*\xs cm]
       \xlabel{(c)}
       \gprime{1}
      \end{scope}
      
      \begin{scope}[xshift=8*\xs cm]
       \xlabel{(d)}
       \gprime{1}
       \node at (x1out)[xnodex]{};
       \foreach \x in {s,a,b}{\node at (\x1out)[xnodex]{};}
       \foreach \x in {a,b,t}{\node at (\x1in)[xnodex]{};}
      \end{scope}
      
      \begin{scope}[xshift=11*\xs cm]
       \xlabel{(e)}
       \gprime{1}
       \node at (y1out)[xnodey]{};
       \foreach \x in {s,a,b}{\node at (\x)[xnodey]{};\node at (\x1out)[xnodey]{};}
       \foreach \x in {s,a,b}{\node at (\x)[xnodey]{};\node at (\x1out)[xnodey]{};}
       \foreach \x/\y in {y1out/s1out,s/s1out,y1out/a1out,a/a1out,y1out/b1out,b/b1out}{\draw[xedgex] (\x) to (\y);}
      \end{scope}
      
    \end{tikzpicture}
  \caption{Illustration to~\cref{constr:1rgbpNPhconsthab-A,constr:1rgbpNPhconsthab-B}.
	  Part (a) shows an exemplary directed graph which is a \yes-instance for \prob{DHP}.
	  Applying \cref{constr:1rgbpNPhconsthab-A} on (a) yields (b).
	  Applying \cref{constr:1rgbpNPhconsthab-B} on (b) yields  the instance whose graph is depicted in (c) and two habitats of which are depicted in (d) and (e).
	  Vertices marked yellow in (d) are contained in the habitat $X_{\rm out}$.
	  Vertices marked red in (e) are contained in the habitat~$Y_{\rm out}$.
	  The graph induced by~$Y_{\rm out}$ contains the red edges.
  }
  \label{fig:1rgbp-paranp}
\end{figure}

As the habitats~$X^*_{\rm out}$, $X^*_{\rm in}$, $Y_{\rm out}$, and $Y_{\rm in}$ induce trees in~$G$, we have the following.

\begin{observation}
	\label{obs:1rgbpNPhconsthab}
	If $(G,\calH,k)$,
	obtained from~$(G',s,t)$ using~\cref{constr:1rgbpNPhconsthab-B},
	is a \yes-instance of~ \RgbpAcr[1]{},
	then every solution contains all edges contained in~$P_v$ for every~$v\in W$
	and all edges incident with~$x_{\rm out}$,
	with~$x_{\rm in}$,
	with $y_{\rm out}$,
	and with~$y_{\rm in}$.
\end{observation}

\begin{lemma}
	\label{lem:1rgbpNPhconsthab}
	Instance $\I'=(G',s,t)$,
	obtained from~$(D,s,t)$ using~\cref{constr:1rgbpNPhconsthab-A},
	is a \yes-instance of \prob{HP}
	if and only if
	$\I=(G,\calH,k)$,
	obtained from~$(G',s,t)$ using~\cref{constr:1rgbpNPhconsthab-B},
	is a \yes-instance of~ \RgbpAcr[1]{}.
\end{lemma}

\begin{proof}
	Let~$F'\ceq \bigcup_{v\in W} E(P_v) \cup \{e\in E\mid e\cap \{x_{\rm out},x_{\rm in},y_{\rm out},y_{\rm in}\} \neq \emptyset\}$ and let~$n \ceq |W|$.
	Note that~$|F'| = 2(n-2) + 2 + 4(n-1)$.

	\RD{}
	Let~$P = (s, v^2, \dots, v^{n-2}, t)$ be an~$s$-$t$ path in~$D$ that visits every vertex exactly once.
	We claim that $F \ceq F' \cup F''$ is a solution for instance~$\I$, where $F'' \ceq \{u_{\rm out}, v_{\rm in} \mid (u, v) \in A(P)\}$.
	Clearly $|F| \le |F'| + |F''| = k$.
	Note that the set $F'$ already connects the habitats~$X^*_{\rm out}$, $X^*_{\rm in}$, $Y_{\rm out}$, and $Y_{\rm in}$.
	Note that $P$ is a subgraph of~$D$ that is weakly connected and in which every vertex has indegree one and every vertex has outdegree one, except for~$s$ (indegree zero) and~$t$ (outdegree zero).
	Hence, for every~$v \in W$ there exists an edge~$(v, w) \in A(P)$, and thus~$\{v_{\rm out}, w_{\rm in}\} \in F''$.
	Therefore, $G[F][X_{\rm in}]$ is connected.
	The argumentation for $G[F][X_{\rm out}]$ being connected is analogous.
	Finally, as $P$ is a connected subgraph, $F$ contains the edges of an $s$-$t$ path that contains all vertices in~$V_{\rm all}$.

	\LD{}
	Let~$F$ be a solution to~$\I$.
	Due to~\cref{obs:1rgbpNPhconsthab},
	we know that~$F'\subseteq F$,
	and hence for~$F''\ceq F\setminus F'$ we have~$|F''|\leq k-(6(n-1)) = n-1$.
	By definition of~$X_{\rm out}$ and~$X_{\rm in}$,
	we know that in~$G[F]$,
	every~$v_{\rm out}$ is adjacent to at least one~$w_{\rm in}$,
	and every~$v_{\rm in}$ is adjacent to at least one~$w_{\rm out}$.
	Thus,
	in the graph~$P\ceq (W,E^*)$ with~$E^*=\{(v,w)\mid (v_{\rm out},w_{\rm in})\in F''\}$,
	every vertex has indegree and outdegree one, except for~$s$ (indegree zero) and~$t$ (outdegree zero).
	We claim that $P$ is weakly connected.
	Consider any two vertices $v, w \in W$.
	By our assumption there exists a $v$-$w$ path $P'$ in $G[F][V_{\rm all}]$.
	Note that~$G[F'][V_{\rm all}]$ has~$n$ connected components, each of which contains exactly one vertex in~$W$.
	Hence, $P'$ contains at least one edge in $F''$, and an additional edge in $F''$ for each additional vertex in $W$ that is visited by $P'$.
	This edge set $E(P') \cap F''$ corresponds to the edges of an undirected $v$-$w$ path in $P$.
	Hence, $P$ is connected.
	Together with the above properties of $P$, it follows that $P$ is a Hamiltonian~$s$-$t$ path.
\end{proof}

Lastly, 
we show that \RgbpAcr[1]{} becomes tractable for $r=2$.
Let~$\alpha \colon \N \to \N$ be the inverse of the single-valued Ackermann function.

\begin{proposition}
	\RgbpAcr[1]{} is solvable in $\O((n+m) \alpha(n))$ time if $r=2$.
\end{proposition}
\begin{proof}
	Assume that both~$G[V_1]$ and~$G[V_2]$ are connected
	(otherwise,
	safely return \no)
	and that~$V_1 \cap V_2 \ne \emptyset$
	(otherwise, 
	a tree spanning over~$V_i$ for each~$i \in \{1, 2\}$ is a valid, minimum-size solution).
	We first compute a spanning forest~$T_{\cap}$ within~$G[V_1 \cap V_2]$, using breadth-first search.
	Afterwards, for each~$i \in \{1, 2\}$,
	we run Kruskal's~\cite{Kruskal56} algorithm to extend the forest~$T_{\cap}[V_i]$ to a spanning tree~$T_i$ that spans over the vertices in~$V_i$.
	Let~$F \ceq E(T_1) \cup E(T_2)$.
	We return \yes{} if and only if~$|F| \le k$.
	As each $v \in V_i$ is visited at most once, the algorithm runs in $\O((n+m) \alpha(n))$ time by using the disjoint-set structure~\cite{Tarjan76}.

	To prove the correctness of the algorithm,
	we show that~$F$ is a minimum-cardinality solution.
	Since both~$G[V_1]$ and~$G[V_2]$ are connected, 
	$G[F][V_i]$ is connected for each~$i \in \{1, 2\}$.
	It remains to show that~$F$ is of minimum cardinality.
	Consider some minimum-cardinality solution~$F'$.
	Let~$F_i'\ceq E(G[F'][V_i])$ for each~$i\in\{1,2\}$,
	and let~$F_\cap'\ceq E(G[F'][V_1\cap V_2])$.
	Observe that~$|E(T_{\cap})|\geq |F_\cap'|$ as otherwise there is cycle in~$G[F'][V_1\cap V_2]$ contradicting the fact that~$F'$ is of minimum-cardinality.
	It follows that
	\begin{flalign*}
		&&|F'| = |F_1'| + |F_2'| - |F_\cap|
        &\geq |V_1|-1 + |V_2|-1 - |F_\cap| \\
	&&&\geq |V_1|-1 + |V_2|-1 - |E(T_\cap)| = |F|. &\squareforqed
	\end{flalign*}
\end{proof}

\subsection{One hop between habitat patches (\texorpdfstring{$d=2$}{d=2})}
\label{ssec:2rgbp}

In this section we prove that~\RgbpAcr[2]{}
is already~\NP-complete even if there are two habitats and the graph has maximum degree four,
or if there is only one habitat.
Afterwards we show that \RgbpAcr[2]{} still admits a problem kernel with respect to $k$.
If the graph is planar, we can show that the kernelization is polynomial in the number of vertices.

\begin{proposition}
 \label{prop:2rgbp}
 \RgbpAcr{}  with~$d\geq 2$ is \NP-complete 
 even if
 (i) $r=2$ and~$\Delta\leq 4$
 or
 (ii) $r=1$ and the input graph has diameter~$2d$.
\end{proposition}

For the sake of presentation,
we prove~\cref{prop:2rgbp}(i) for~$d=2$.
Afterwards,
we briefly explain how to adapt the proof for~$d>2$ and for \cref{prop:2rgbp}(ii).

\begin{construction}
 \label{constr:2rgbp}
 Let~$\I=(G, k)$ be an instance of \prob{3-Regular Vertex Cover} with~$G=(V, E)$ and~$V=\set{n}$
 construct an instance of~\RgbpAcr[2]{}
 with graph~$G'=(V',E')$, 
 habitat sets~$V_1$ and~$V_2$,
 and integer~$k'\ceq |E|+(n-1)+k$ as follows 
 (see~\cref{fig:2rgbp}(a) for an illustration).
 \begin{figure}[t]
  \centering
  \begin{tikzpicture}

    \def\xr{0.77}
    \def\yr{0.7}
    \def\ys{1.5*\yr}
    \def\teps{0.125}
    \tikzstyle{xnode}=[circle,fill,scale=0.6,draw,color=\thecolor]
    \tikzstyle{xnodef}=[circle,scale=0.6,draw,color=\thecolor,fill=white]
    \tikzstyle{xnodex}=[label={[xshift=-0.75*\xr cm]0:$\cdots$},color=\thecolor,inner sep=6pt]
    \tikzstyle{xedge}=[thick,-,color=\thecolor]

    \node at (-4.25*\xr,1*\yr)[]{(a)};
    
    \foreach \x/\y in {1/xnode,2/xnodex,3/xnode,4/xnodex,5/xnode,6/xnodex,7/xnode,8/xnodex,9/xnode}{
    \node (e\x) at (\x*\xr-5*\xr,0)[\y]{};
    }
    \draw[dotted,rounded corners,draw] ($(e1.south west)-(\teps,\teps)$) rectangle ($(e9.north east)+(\teps,\teps)$);
    \foreach \x/\y in {1/xnodef,2/xnodex,3/xnodef,4/xnodex,5/xnodef,6/xnodex,7/xnodef}{
    \node (v\x) at (\x*\xr-4*\xr,-1*\ys)[\y]{};
    }
    \draw[dotted,rounded corners,draw] ($(v1.south west)-(\teps,\teps)$) rectangle ($(v7.north east)+(\teps,\teps)$);
    \foreach \x/\y in {1/xnode,2/xnodex,3/xnode,4/xnodex,5/xnode,6/xnodex,7/xnode}{
    \node (z\x) at (\x*\xr-4*\xr,-2*\ys)[\y]{};
    }
    \draw[dotted,rounded corners,draw] ($(z1.south west)-(\teps,\teps)$) rectangle ($(z7.north east)+(\teps,\teps)$);
    \foreach \x/\y in {3/3,3/5,5/3,5/2,7/5,7/6}{
      \draw[xedge] (e\x) to (v\y);
    }
    \foreach \x/\y in {2/1,8/7}{
      \draw[draw=none] (e\x) to node[midway]{$\vdots$}(v\y);
    }
    \foreach \x in {1,...,7}{
      \draw[xedge] (z\x) to (v\x);
    }
    \foreach \x/\y in {1/2,2/3,3/4,4/5,5/6,6/7}{
      \draw[xedge] (z\x) to (z\y);
    }
    
    \contourlength{0.09em}
    \newcommand{\ltb}[3]{
      \node at (#1)[label={[align=center,font=\footnotesize,color=black]#2},label={[align=center,font=\scriptsize,color=black]-90:\contour*{white}{#3}}]{};
    }
    
    \ltb{e3}{\footnotesize$e=$\\\footnotesize$\{i,j\}$}{$\in V_1$};
    \ltb{e5}{\footnotesize$e'=$\\\footnotesize$\{i,j'\}$}{$\in V_1$};
    \ltb{e7}{\footnotesize$e''=$\\\footnotesize$\{i',j\}$}{$\in V_1$};
    
    \ltb{z3}{\contour*{white}{$x_i$}}{$\in V_1,V_2$};
    \ltb{z5}{\contour*{white}{$x_j$}}{$\in V_1,V_2$};
  
    \ltb{v3}{$v_i$}{};
    \ltb{v5}{$v_j$}{};
  \end{tikzpicture}
  \hfill
  \begin{tikzpicture}

    \def\xr{0.77}
    \def\yr{0.7}
    \def\ys{1.5*\yr}
    \def\teps{0.125}
    \tikzstyle{xnode}=[circle,fill,scale=0.6,draw,color=\thecolor]
    \tikzstyle{xnodef}=[circle,scale=0.6,draw,color=\thecolor,fill=white]
    \tikzstyle{xnodex}=[label={[xshift=-0.64*\xr cm]0:$\cdots$},color=\thecolor]
    \tikzstyle{xedge}=[thick,-,color=\thecolor]

    \node at (-4.25*\xr,1*\yr)[]{(b)};
    
    \foreach \x/\y in {1/xnode,2/xnodex,3/xnode,4/xnodex,5/xnode,6/xnodex,7/xnode,8/xnodex,9/xnode}{
    \node (e\x) at (\x*\xr-5*\xr,0)[\y]{};
    }
    \draw[dotted,rounded corners,draw] ($(e1.south west)-(\teps,\teps)$) rectangle ($(e9.north east)+(\teps,\teps)$);
    \foreach \x/\y in {1/xnodef,2/xnodex,3/xnodef,4/xnodex,5/xnodef,6/xnodex,7/xnodef}{
    \node (v\x) at (\x*\xr-4*\xr,-1*\ys)[\y]{};
    }
    \draw[dotted,rounded corners,draw] ($(v1.south west)-(\teps,\teps)$) rectangle ($(v7.north east)+(\teps,\teps)$);
    \node (x) at (0*\xr,-2*\ys)[xnodef]{};

    \foreach \x/\y in {3/3,3/5,5/3,5/2,7/5,7/6}{
      \draw[xedge] (e\x) to (v\y);
    }
    \foreach \x/\y in {2/1,8/7}{
      \draw[draw=none] (e\x) to node[midway]{$\vdots$}(v\y);
    }
    \foreach \x in {1,...,7}{
      \draw[xedge] (x) to (v\x);
    }
    
    \contourlength{0.09em}
    \newcommand{\ltb}[3]{
      \node at (#1)[label={[align=center,font=\footnotesize,color=black]#2},label={[align=center,font=\scriptsize,color=black]-90:\contour*{white}{#3}}]{};
    }
    
    \ltb{e3}{\footnotesize$e=$\\\footnotesize$\{i,j\}$}{$\in V_1$};
    \ltb{e5}{\footnotesize$e'=$\\\footnotesize$\{i,j'\}$}{$\in V_1$};
    \ltb{e7}{\footnotesize$e''=$\\\footnotesize$\{i',j\}$}{$\in V_1$};
    
    \ltb{x}{\contour*{white}{$x$}}{$\in V_1$};
  
    \ltb{v3}{$v_i$}{};
    \ltb{v5}{$v_j$}{};
  \end{tikzpicture}
  \caption{Illustration for~\RgbpAcr[2]{} with 
  (a) $r=2$ and~$\Delta=4$ ($k'=m+(n-1)+k$) and
  (b) $r=1$ ($k'=m+k$).}
  \label{fig:2rgbp}
\end{figure}

 Add the vertex set~$V_E\ceq \{v_e\mid e\in E\}$
 and add~$v_e$ with $e=\{i,j\}\in E$ to habitat~$V_1$.
 Next,
 add the vertex sets~$V_G=\{v_i\mid i\in V\}$,
 and connect each~$v_i$ with all edge-vertices corresponding to an edge incident with~$i$,
 i.e.,
 add the edge set~$E_G\ceq \bigcup_{i\in V}\{\{v_i,v_e\}\mid i\in e\}$.
 Next,
 add the vertex set~$V_X\ceq \{x_i\mid i\in V\}$,
 connect each~$x_i$ with~$v_i$,
 and add~$x_i$ to~$V_1$ and to~$V_2$.
 Finally,
 add the edge set~$\{\{x_i,x_{i+1}\}\mid i\in\set{n-1}\}$.
\cqed
\end{construction}

\begin{observation}
 \label{obs:2rgbp}
 Let~$\I=(G,k)$ be an instance of \prob{3-Regular Vertex Cover} and let~$\I'=(G',\{V_1,V_2\},k')$ be the instance obtained from~$\I$ using~\cref{constr:2rgbp}.
 If~$\I'$ is a \yes-instance,
 then every solution contains all edges in~$G[V_X]$.
\end{observation}

\begin{proof}
 Suppose not,
 and let~$F$ be a solution without some edge~$\{x_i,x_{i+1}\}$.
 Note that in~$G-\{\{x_i, x_{i+1}\}\}$, the distance between~$x_i$ and~$x_{i+1}$ is at least four;
 thus~$G[F]^2[V_X] = G[F]^2[V_2]$ is not be connected.
 A contradiction.
 \lqed
\end{proof}

\begin{lemma}
 \label{lem:2rgbp:edinc}
 Let~$\I=(G,k)$ be an instance of \prob{3-Regular Vertex Cover} and let~$\I'=(G',f,k')$ be the instance obtained from~$\I$ using~\cref{constr:2rgbp}.
 If~$\I'$ is a \yes-instance,
 then there is a solution~$F\subseteq E(G')$ such that~$\deg_{G'[F]}(v_e)=1$ for all~$e\in E(G)$.
\end{lemma}

\begin{proof}
 Clearly, in every solution,
 we have~$\deg_{G'[F]}(v_e)\geq 1$.
 Let~$F$ be a minimum solution with a minimum number of edges incident to vertices in~$\{v_e \mid e\in E\}$.
 Suppose that there is at least one~$e=\{i,j\}\in E$ such that~$\deg_{G'[F]}(v_e)=2$, that is, $\{v_e, v_i\}, \{v_e, v_j\} \in F$.
 Since~$F$ is a solution, there is a path~$P$ in~$G'[F]$ from~$v_e$ to some~$x_i$.
 Let~$\{v_e,v_i\}$ be the first edge on this path.
 Let~$F'\ceq (F \setminus\{v_e,v_j\})\cup\{v_j,x_j\}$.
 We claim that~$F'$ is a solution,
 yielding a contradiction to the fact that~$F$ is a solution with a minimum number of edges incident with vertices in~$V_E$.

 Only a vertex~$v_{e'}$ can be disconnected from any~$V_X$ by removing~$\{v_e,v_j\}$ from~$F$.
 This vertex cannot be on the path~$P$,
 and hence is connected to~$v_e$ via edge~$\{v_e,v_j\}$.
 Since now edge~$\{v_j,x_j\}$ is present,
 $v_{e'}$ is again connected to~$V_X$.
 \lqed
\end{proof}

\begin{lemma}
 \label{lem:2rgbp:cor}
 Let~$\I=(G,k)$ be an instance of \prob{3-Regular Vertex Cover} and let~$\I'=(G',\{V_1,V_2\},k')$ be the instance obtained from~$\I$ using~\cref{constr:2rgbp}.
 Then~$\I$ is a \yes-instance if and only if~$\I'$ is a \yes-instance.
\end{lemma}

\begin{proof}
 \RD{}
 Let~$S\subseteq V$ be a vertex cover of size~$k$ in~$G$.
 We construct a solution~$F\subseteq E'$ as follows.
 Let~$F_X=\bigcup_{i=1}^{n-1} \{\{x_i,x_{i+1}\}\}$
 and~$F_V\ceq \{\{v_i,x_i\}\mid i\in S\}$.
 We define the auxiliary function~$g\colon E\to V'$ with~$g(\{i,j\})=v_{\min(\{i,j\}\cap S)}$.
 Let~$F_E\ceq \bigcup_{e=\{i,j\}\in E} \{v_e,g(e)\}$.
 Let~$F\ceq F_X\cup F_V\cup F_E$. 
 Note that~$|F|=|F_X|+|F_V|+|F_E|\leq |E|+(n-1)+k = k'$.
 Moreover, 
 every~$v_e\in V_E$ is connected to~$x_i$ via a path~$(v_e,v_i,x_i)$, 
 where~$i\in (e\cap S)$.
 Finally,
 observe that~$G'[F][V_X]$ is connected.
 
 \LD{}
 Let~$\I'$ be a \yes-instance.
 Due to \cref{lem:2rgbp:edinc} there
 is a solution~$F\subseteq E'$ such that~$\deg_{G'[F]}(v_e)=1$ for all~$e\in E$.
 Due to~\cref{obs:2rgbp},
 we know that the edges $\bigcup_{i=1}^{n-1} \{\{x_i,x_{i+1}\}\}\subseteq F$.
 Let~$S\ceq \{i\in V\mid \{v_i,x_i\}\in F\}$.
 We claim that~$S$ is a vertex cover.
 Suppose not,
 that is,
 there is an edge~$e\in E$ such that~$e\cap S=\emptyset$.
 That means that the unique neighbor of~$v_e$,
 say~$v_i$,
 is not adjacent with~$x_i$ in~$G'[F]$.
 Since $\deg_{G'[F]}(v_e)=1$ for all~$e\in E$,
 $N_{G'[F]}[v_i]$ forms a connected component in~$G'[F]^2$ not containing~$x_i$.
 This contradicts the fact that
 $F$ is a solution.
 \lqed
\end{proof}

\begin{remark}
  \begin{inparaenum}[(i)]
  \item To make the reduction work for~$d\geq 3$,
  it is enough to subdivide each edge~$\{v_e,v_i\}$
  $(d-2)$ times and set~$k'\ceq (d-1)m+(n-1)+k$.
  \item If we contract all~$x_i$,
  set~$V_2=\emptyset$ 
  (i.e., only one habitat remains),
  and set~$k'\ceq (d-1)m+k$,
  then the reduction is still valid
  (see~\cref{fig:2rgbp}(b) for an illustration).
  \end{inparaenum}
  Thus,
  \cref{prop:2rgbp}(ii) follows.
\end{remark}

The reduction in the proof of~\cref{prop:2rgbp} 
requires $k$ to be linear in the input instance's size.
We next prove that, indeed, \RgbpAcr[2]{} is fixed-parameter tractable with respect to~$k$
by showing that it admits a problem kernel of size exponential in~$k$.
\begin{proposition}
 \label{prop:rgbp:kernel}
 \RgbpAcr[2]{} admits a problem kernel with at most $2k+\binom{2k}{k}$ vertices,
 at most $\binom{2k}{2}+k\binom{2k}{k}$ edges,
 and at most~$2^{2k}$ habitats.
\end{proposition}

Let~$\bar{V}\ceq V\setminus \bigcup_{V'\in\calH} V'$ for a graph~$G=(V,E)$ and habitat set~$\calH=\{V_1,\dots,V_r\}$.
The following reduction rules are immediate.

\begin{rrule}
 \label{rr:immediate}
 \begin{inparaenum}[(i)]
  \item If~$|V_i|=1$ for some~$i$,
  delete~$V_i$.
  \item If a vertex in~$\bar{V}$ is of degree at most one,
  delete it.
  \item If there is an~$i\in\set{r}$ with~$|V_i|>1$ and an~$v\in V_i$ of degree zero,
  return a trivial \no-instance.
  \item If there is a vertex~$v\in V\setminus\bar{V}$ of degree at most one,
  delete it (also from~$V_1,\dots,V_r$),
  and set~$k\ceq k-1$.
 \end{inparaenum}
\end{rrule}
Clearly, 
$k$ edges can connect at most~$2k$ vertices;
thus we obtain the following.

\begin{rrule}
 \label{rr:few-habitat-vertices}
 If~$|V\setminus\bar{V}|>2k$,
 then return a trivial \no{}-instance.
\end{rrule}

So we have at most~$2k$ vertices in habitats.
Next, we upper-bound the number of non-habitat vertices.
No minimal solution has edges between two such vertices.

\begin{rrule}
 \label{rr:no-bar-edges}
 If there is an edge~$e \in E$ with $e\subseteq \bar{V}$,
 then delete~$e$.
\end{rrule}

Moreover,
no minimum solution connects through non-habitat twins.

\begin{rrule}
 \label{rr:no-twins}
 If~$N(v)\subseteq N(w)$ for distinct~$v,w\in \bar{V}$,
 then delete~$v$.
\end{rrule}

We still need to bound the number of vertices in~$\bar V$.
For an $n$-element set~$S$ let~$\mathcal F \subseteq 2^{S}$ be a family of subsets such that
for every~$A, B \in \mathcal F$ we have~$A \not\subseteq B$.
Then~$|\mathcal F| \le \binom{n}{\lfloor n/2 \rfloor}$ by Sperner's Theorem.
Hence,
after 
applying the reduction rules,
we get an instance with at most~$2k+\binom{2k}{k}$ vertices 
and~$\binom{2k}{2}+2k\binom{2k}{k}$ edges.

Finally, 
we can upper-bound the number of habitats
by simply deleting duplicates.

\begin{rrule}
\label{rr:habitats}
If $V_i=V_j$ for distinct~$i,j\in\set{r}$,
then delete~$V_j$.
\end{rrule}

\noindent
It follows that we can safely assume that~$r\leq 2^{2k}$.
Thus,
\cref{prop:rgbp:kernel} follows.
Unfortunately,
improving the problem kernel 
to polynomial-size is unlikely.

\begin{proposition}
 \label{prop:rgbp:nopk}
 Unless~\NPincoNPslashpoly,
 \RgbpAcr{} for~$d\geq 2$ admits no problem kernel of size~$k^{\O(1)}$,
 even if~$r\geq 1$ is constant.
\end{proposition}

\noindent
We will give a linear parametric transformation from the following problem:

\decprob{Set Cover (SC)}{setcover}
{A universe~$U$, a set~$\calF\subseteq 2^U$ of subsets of~$U$, and an integer~$k$.}
{Is there~$\calF'\subset\calF$ with~$|\calF'|\leq k$ such that~$\bigcup_{F\in\calF'} F=U$?}

\noindent
The construction is basically the same as for \cref{prop:2rgbp}(ii).
Note that \prob{Set Cover} admits no problem kernel of size polynomial in~$|U|+k$,
unless~$\NPincoNPslashpoly$~\cite{DomLS14}.

\begin{proof}
 Let~$\I=(U,\calF,k)$ be an instance of~\prob{Set Cover},
 with~$U=\{u_1,\dots,u_n\}$.
 Construct an instance~$\I'\ceq (G,V_1,k')$ of \RgbpAcr[2]{} with~$k'=|U|+k$ as follows 
 (see~\cref{fig:2rgbp:nopk}).
 \begin{figure}[t]
  \centering
  \begin{tikzpicture}

    \def\xr{1.1}
    \def\yr{1}
    \def\ys{1.5*\yr}
    \def\teps{0.125}
    \tikzstyle{xnode}=[circle,fill,scale=0.6,draw,color=\thecolor]
    \tikzstyle{xnodef}=[circle,scale=0.6,draw,color=\thecolor,fill=white]
    \tikzstyle{xnodex}=[label={[xshift=-0.4*\xr cm]0:$\cdots$},color=\thecolor]
    \tikzstyle{xedge}=[thick,-,color=\thecolor]

    \foreach \x/\y in {1/xnode,2/xnodex,3/xnode,4/xnodex,5/xnode,6/xnodex,7/xnode,8/xnodex,9/xnode}{
    \node (e\x) at (\x*\xr-5*\xr,-1*\ys)[\y]{};
    }
    \draw[dotted,rounded corners,draw] ($(e1.south west)-(\teps,\teps)$) rectangle ($(e9.north east)+(\teps,\teps)$);
    \foreach \x/\y in {1/xnodef,2/xnodex,3/xnodef,4/xnodex,5/xnodef,6/xnodex,7/xnodef}{
    \node (v\x) at (\x*\xr-4*\xr,0*\ys)[\y]{};
    }
    \draw[dotted,rounded corners,draw] ($(v1.south west)-(\teps,\teps)$) rectangle ($(v7.north east)+(\teps,\teps)$);
    \node (x) at (0*\xr,-2*\ys)[xnodef]{};

    \foreach \x/\y in {3/2,3/3,3/5,5/1,5/3,5/5,5/7,7/4,7/5,7/6}{
      \draw[xedge] (e\x) to (v\y);
    }
    \foreach \x/\y in {2/1,8/7}{
      \draw[draw=none] (e\x) to node[midway]{$\vdots$}(v\y);
    }
    \foreach \x in {1,...,9}{
      \draw[xedge] (x) to (e\x);
    }
    
    \contourlength{0.09em}
    \newcommand{\ltb}[3]{
      \node at (#1)[label={[align=center,font=\footnotesize,color=black]\contour*{white}{#2}},label={[align=center,font=\scriptsize,color=black]-90:\contour*{white}{#3}}]{};
    }
    
    \ltb{e3}{$v_{F'}$}{};
    \ltb{e5}{$v_F$}{};
    \ltb{e7}{$v_{F''}$}{};
    
    \ltb{x}{$x$}{$\in V_1$};
  
    \ltb{v1}{$u_1$}{$\in V_1$};
    \ltb{v7}{$u_n$}{$\in V_1$};
    \ltb{v3}{$u_i$}{$\in V_1$};
    \ltb{v5}{$u_j$}{$\in V_1$};
    
    \node[right =of v7,xshift=-\xr*0.5 cm]{$V_U$};
    \node[right =of e9,xshift=-\xr*0.5 cm]{$V_\calF$};
  \end{tikzpicture}
  \caption{Illustration for the construction in the proof of~\cref{prop:rgbp:nopk} for~\RgbpAcr[2]{} with~$r=1$. 
  In this example, 
  $U=\{u_1,\dots,u_n\}$ and
  we have $\{u_1,u_i,u_j,u_n\}= F\in\calF$.}
  \label{fig:2rgbp:nopk}
  \end{figure}
 Let~$G$ be initially empty.
 Add the vertex set~$V_U\ceq U$,
 the vertex set~$V_\calF\ceq \{v_F\mid F\in\calF\}$,
 and the vertex~$x$.
 Set~$V_1\ceq V_U\cup\{x\}$.
 Make each vertex in~$V_\calF$ adjacent with~$x$.
 Finally,
 for each~$F\in\calF$,
 add the edge set~$\{\{v_i,v_F\}\mid u_i\in F\}$.
 
 The proof that~$\I$ is a \yes-instance if and only if~$\I'$ is a \yes-instance 
 is analogous with the correctness proof for \cref{prop:2rgbp}(ii).
 
 Since \prob{Set Cover} admits no problem kernel of size polynomial in~$|U|+k$,
 unless~$\NPincoNPslashpoly$~\cite{DomLS14},
 neither does~\RgbpAcr[2]{} when parameterized by~$k'=|U|+k$.
 \lqed
\end{proof}

\cref{prop:rgbp:nopk} holds for general graphs.
In fact, 
for planar graphs,
the above reduction rules allow for 
an~$\O(k^3)$-vertex kernel.
The number of habitats in the kernel however may still be exponential in $k$.

\begin{proposition}
	\label{prop:rgbp:planar}	
	\RgbpAcr[2]{} on planar graphs admits a problem kernel with~$\O(k^3)$ vertices and edges and at most~$2^{2k}$ habitats.
\end{proposition}

\begin{observation}
  \label{obs:2rgbpplanar}
	Suppose all reduction rules were applied exhaustively.
	Then
	\begin{enumerate}[(i)]
	\item there are at most~$\binom{2k}{2}$ vertices of degree two in~$\bar V$, and
	\item there are at most~$3\binom{2k}{3}$ vertices of degree at least three in~$\bar V$.
	\end{enumerate}
\end{observation}

\begin{proof}
	\begin{inparaenum}[\itshape (i)]
	\item By \cref{rr:no-bar-edges,rr:few-habitat-vertices,rr:no-twins},
		every degree-two vertex in~$\bar V$ has a pairwise different pair of neighbors in~$V \setminus \bar V$.
		As there are~$\binom{2k}{2}$ (unordered) vertex pairs in $V \setminus \bar V$,
		there are at most $\binom{2k}{2}$ degree-two vertices in~$\bar V$,
		otherwise one of the reduction rules was not applied exhaustively.

	\item Any three vertices~$u,v,w$ in a planar graph share at most two neighbors,
		that is,
		$|N(u)\cap N(v)\cap N(w)| \le 2$.
		Suppose there are more than~$3\binom{2k}{3}$ vertices in~$\bar V$ of degree at least three.
		Then,
		by \cref{rr:no-bar-edges,rr:few-habitat-vertices,rr:no-twins},
		there are three vertices~$u,v,w\in\bar V$ such that~$|N(u)\cap N(v)\cap N(w)| \ge 3$,
		a contradiction to~$G$ being planar.
	\end{inparaenum}
\end{proof}

As~$|V\setminus \bar V| \le 2k$ and we deleted all degree-one vertices,
\cref{prop:rgbp:planar} follows.

\subsection{At least two hops between habitat patches (\texorpdfstring{$d\ge3$}{d≥3})}
\label{ssec:3rgbp}

If the data is more sparse,
that is,
the observed habitats to connect are rather scattered,
then the problem becomes significantly harder to solve from the parameterized complexity point of view.

\begin{proposition}
 \label{prop:3rgbp}
 \RgbpAcr{}  with~$d\geq 3$ is \NP-complete and 
 \W{1}-hard when parameterized by~$k+r$.
\end{proposition}

We give the construction for~$d$ being odd.
Afterwards,
we explain how to adapt the reduction to~$d$ being even.
The reduction is from the \prob{Multicolored Clique} problem,
where,
given a $k$-partite graph~$G=(U^1,\dots,U^k,E)$,
the question is whether there is a clique containing exactly one vertex from each part.
\prob{Multicolored Clique} is \NP-hard and~\W{1}-hard when parameterized by~$k$.

\begin{construction}
 \label{constr:3rgbp}
 Let~$(G)$ with~$G=(U^1,\dots,U^k,E)$ be an instance of \prob{Multicolored Clique}
 where~$G[U^i]$ forms an independent set for every~$i\in\set{k}$.
 Assume without loss of generality that~$U^i=\{u^i_1,\dots,u^i_{|V^i|}\}$.
 Let~$k'\ceq \frac{(d-1)}{2}k+\binom{k}{2}$.
 Construct the instance~$(G',\{V_1,\dots,V_{\binom{k}{2}}\},k')$ as follows
 (see~\cref{fig:3rgbp} for an illustration).
 \begin{figure}[t]
  \centering
    \begin{tikzpicture}

      \usetikzlibrary{calc}
      \def\xr{1}
      \def\yr{0.775}
      \def\teps{0.175}
      \tikzstyle{xnode}=[circle,fill,scale=0.6,draw,color=\thecolor]
      \tikzstyle{xedge}=[thick,-,color=\thecolor]
      \tikzstyle{xnodef}=[circle,scale=0.6,draw,color=\thecolor,fill=white]

      \newcommand{\xpart}[6]{
        \begin{scope}[xshift=#5*\xr cm,yshift=#6*\yr cm,rotate around={#4:(0,0)}]
          \node (u#1)  at (0,2*\yr)[xnode]{};
          \node (v#11)  at (-1.5*\xr,0*\yr)[xnodef]{};
          \node (v#12)  at (-0.75*\xr,0*\yr)[rotate=#4]{$\cdots$};
          \node (v#13)  at (0*\xr,0*\yr)[xnodef]{};
          \node (v#14)  at (0.75*\xr,0*\yr)[rotate=#4]{$\cdots$};
          \node (v#15)  at (1.5*\xr,0*\yr)[xnodef]{};

          \foreach \x in {1,...,5}{
            \draw[xedge] (u#1) to (v#1\x);
            \draw[dash pattern=on \pgflinewidth off 8pt, fill=white,thick, line width=3pt,color=\thecolor] (u#1) to (v#1\x);
          }
          \draw[densely dotted,rounded corners] ($(v#11)-(\teps,\teps)$) rectangle ($(v#15)+(\teps,\teps)$);
        \end{scope}
        \node at (v#15)[label={[]#3}]{};
        \node at (u#1)[label={[]#2}]{};
      }
      
      \xpart{1}{90:$\in V_\ell$ if~$i\in g^{-1}(\ell)$}{90:$U^i$}{45}{-3}{2};
      \xpart{2}{-90:$\in V_\ell$ if~$j \in g^{-1}(\ell)$}{180:$U^j$}{135}{-3}{-2};
      \xpart{3}{-90:$\in V_\ell$ if~$j' \in g^{-1}(\ell)$}{-90:$U^{j'}$}{225}{3}{-2};
      \xpart{4}{90:$\in V_\ell$ if~$i' \in g^{-1}(\ell)$}{0:$U^{i'}$}{315}{3}{2};

      \node (top) at (0,3.2*\yr)[scale=2]{$\cdots$};
      \node (bot) at (0,-3.2*\yr)[scale=2]{$\cdots$};
      \draw[xedge] (v13) to (v25) to (v33) to (v45) to (v13) to (v33);
      \draw[xedge] (v25) to (v45);
      \foreach \x/\y in {1/3,2/5,3/3,4/5}{
        \draw[xedge] (v\x\y) to (top);
        \draw[xedge] (v\x\y) to (bot);
      }
      \draw[draw=none] (v11) to node[midway,below,sloped,color=black!80]{\small $(d-1)/2$ edges}(u1);

      \end{tikzpicture}
  \caption{Illustration to~\cref{constr:3rgbp} for~\RgbpAcr{} for~$d\geq 3$.}
  \label{fig:3rgbp}
 \end{figure}

 Let~$g\colon \binom{\set{k}}{2}\to \set{\binom{k}{2}}$ be a bijective function.
 Let~$G'$ be initially~$G$.
 For each~$i\in\set{k}$,
 add a vertex~$v_i$ to~$G'$,
 add~$v_i$ to each habitat~$V_\ell$ with~$i\in g^{-1}(\ell)$,
 and connect~$v_i$ with~$u^i_j$ for each~$j\in\set{u^i_{|U^i|}}$ via a path with~$\frac{d-1}{2}$ edges,
 where~$v_i$ and~$u_i^j$ are the endpoints of the path.
 \cqed
\end{construction}

\begin{remark}
 For every even~$d\geq 4$,
 we can adapt the reduction for~$d-1$:
 At the end of the construction,
 subdivide each edge between two vertices that are in the original graph~$G$.
\end{remark}

\begin{observation}
 \label{obs:3rgbp:smallhabs}
 In the obtained instance,
 for every~$\ell\in\set{\binom{k}{2}}$,
 it holds that, 
 $V_\ell=\{v_i,v_j\}$ where~$\{i,j\}=g^{-1}(\ell)$,
 and for every~$i,j\in\set{k}$,
 $i\neq j$,
 it holds that~$\{\ell'\mid \{v_i,v_j\}\subseteq V_{\ell'}\}=\{\ell\}$ with~$\ell=g(\{i,j\})$.
\end{observation}

\begin{observation}
 \label{obs:3rgbp:exone}
 If the obtained instance is a \yes-instance,
 then in every minimal solution~$F$,
 for every~$i\in\set{k}$ there is exactly one~$u^i_j$ in~$G[F]$.
\end{observation}

\begin{proof}
 Note that each~$v_i$ must be connected with at least one vertex from~$U^i$ in~$G'[F]$.
 Thus,
 $|V(G'[F])\cap U^i|\geq 1$.
 Moreover,
 from each~$i,j\in\set{k}$,
 $i\neq j$,
 $F$~must contain an edge between~$U^i$ and~$U^j$,
 since~$\dist_{G'}(v_i,u)+\dist_{G'}(v_j,u')\geq d-1$ for every~$u\in U^i$, 
 $u'\in U^j$.
 Since additionally~$k'= \frac{(d-1)}{2}k+\binom{k}{2}$,
 it follows that~$v_i$ cannot be connected with two vertices from~$U^i$ in~$G'[F][U^i\cup \{v_i\}]$.
 Hence,
 if there are two vertices~$u,u'\in U^i\cap F$,
 with~$u$ being connected to~$v_i$ in~$G'[F][U^i\cup \{v_i\}]$,
 then~$u'$ is not part of an $v_a$-$v_b$~path in~$G'[F]$ of length at most~$d$
 for every~$a,b\in\set{k}$.
 It follows that~$F$ is not minimal.
\end{proof}

\begin{lemma}
 \label{lem:3rgbp}
 Let~$\I=(G)$ with~$G=(U^1,\dots,U^k,E)$ be an instance of \prob{Multicolored Clique} and let~$\I'=(G',\calH,k')$ be the instance obtained from~$\I$ using~\cref{constr:3rgbp}.
 Then~$\I$ is a \yes-instance if and only if~$\I'$ is a \yes-instance.
\end{lemma}

\begin{proof}
 \RD{} 
 Let~$W\subseteq V(G)$ be a multicolored clique.
 Let~$F$ contain~$\binom{W}{2}$ and all edges of a path from~$v_i$ to~$U^i\cap W$.
 We claim that~$F$ is a solution.
 Note that~$|F|=\binom{k}{2}+k\frac{d-1}{2}$.
 Since~$V_\ell$ is of size two for all~$\ell\in\set{\binom{k}{2}}$ (\cref{obs:3rgbp:smallhabs}),
 we only need to show that~$v_i,v_j$ with~$\{i,j\}=g^{-1}(\ell)$
 is connected by a path of length at most~$d$.
 We know that~$v_i$ is connected to some~$u^i_x$ by a path of length~$(d-1)/2$,
 which is adjacent to some~$u^j_y$, 
 which is connected to~$v_j$ by a path of length~$(d-1)/2$.
 Thus, 
 $v_i$ and~$v_j$ are of distance~$d$.
 
 \LD{}
 Let~$F$ be a solution.
 Note that~$|F|=\binom{k}{2}+k\frac{d-1}{2}$.
 We claim that~$W\ceq V(G'[F])\cap V(G)$ is a multicolored clique.
 First, 
 observe that~$|W|=k$ since for every~$v_i$ there is exactly one~$u^i_{\ell_i}$ in~$G'[F]$ (\cref{obs:3rgbp:exone}).
 Suppose that~$W$ is not a multicolored clique,
 that is,
 there are~$U^i$ and~$U^j$ such that there is no edge in~$F$ between them.
 Then~$v_i$ and~$v_j$ are of distance larger than~$d$ in~$G'[F]$,
 contradicting that~$F$ is a solution.
 \lqed
\end{proof}

\section{Connecting Habitats at Short Pairwise Distance}
 \label{sec:cgbp}

In the next problem, 
we require short pairwise reachability.

\decprob{\CgbpTsc{} (\CgbpAcr{})}{cgbp}
{An undirected graph~$G=(V,E)$,
a set~$\mathcal H = \{V_1, \dots, V_r\}$ of habitats where~$V_i\subseteq V$ for all~$i\in\set{r}$,
and~$k \in \Nzero$.}
{Is there a subset~$F\subseteq E$ with~$|F|\leq k$ such that
for every~$i\in\set{r}$
it holds that~$V_i\subseteq V(G[F])$ and~$G[F]^d[V_i]$ is a clique?
}

Note that if $G[F]^d[V_i]$ is a clique, 
then~$\dist_{G[F]}(v,w) \leq d$ for all~$v,w\in V_i$.
Further, \CgbpAcr[2]{} is an unweighted variant of the 2NET problem~\cite{DahlJ04}.

\begin{theorem}
 \CgbpTsc{} is,
 \begin{compactenum}[(i)]
  \item if~$d=1$, linear-time solvable;
  \item if~$d=2$, \NP-hard even on bipartite graphs of diameter three and~$r=1$, 
  and in~\FPT{} regarding~$k$;
  \item if~$d\geq 3$, \NP-hard and~\W{1}-hard regarding~$k$ even if~$r=1$.
 \end{compactenum}
 Further, \CgbpAcr{} is linear-time solvable if the number of habitats and the maximum degree are constant.
\end{theorem}

We first show the linear-time solvability for constant number of habitats and maximum degree.
Afterwards we present the results in (i)-(iii).

\subsection{Graphs of constant maximum degree}

\RgbpAcr[2]{} is \NP-hard if the number~$r$ of habitats and the maximum degree~$\Delta$ are constant 
(\cref{prop:2rgbp}).
\CgbpAcr[2]{} is linear-time solvable in this~case:

\begin{proposition}
 \label{prop:cgbpdelta}
 \CgbpAcr{} admits an~$\O(r\Delta(\Delta-1)^{3d/2})$-sized problem kernel computable in~$\O(r(n+m))$ time.
\end{proposition}

\begin{proof}
	Let~$\I = (G, \calH, k)$ be an instance of \CgbpAcr{}.
	For every~$i\in\set{r}$, fix a vertex~$u_i\in V_i$.
	We assume that we have $V_i \subseteq N_G^d[u_i]$ for all~$i\in\set{r}$, otherwise~$\I$ is a \no-instance.
	Now let~$W_i = N_G^{\lceil 3d/2\rceil}[u_i]$ and let~$G' \ceq G[\bigcup_{i=1}^r W_i]$.
	Note that~$G'$ contains at most~$r\Delta(\Delta - 1)^{\lceil 3d/2\rceil}$ vertices and can be computed by~$r$ breadth-first searches.
	We claim that~$G'$ contains every path of length at most~$d$ between every two vertices~$v, w\in V_i$, for every~$i \in \set{r}$.
	Recall that an edge set~$F \subseteq E$ is a solution if and only if for every~$i\in\set{r}$ and for every~$v, w\in V_i$,
	the graph $G[F]$ contains a path of length at most~$d$ from~$v$ to~$w$.
	As by our claim~$G'$ contains any such path,
	this implies that~$\I$ is a \yes-instance if and only if~$\I'\ceq(G',\calH,k)$ is a \yes-instance (note that~$V_i \subseteq V(G')$ for every~$i \in \set{r}$).

	Assuming that~$V_i \subseteq N_G^d[u_i]$,
	$G[W_i]$ contains all paths of length at most~$d$ between~$u_i$ and any~$v\in V_i$.
	So let~$v, w \in V_i$ be two vertices, both distinct from~$u_i$.
	As $v, w \in N^d_G[u_i]$ and~$W_i=N^{\lceil 3d/2\rceil}_G[u_i]$,
	the subgraph $G[W_i]$ contains all vertices in~$N^{\lceil d/2\rceil}_G[v]$ and~$N^{\lceil d/2\rceil}_G[w]$.
	Consider now a path of length at most~$d$ between~$v$ and~$w$.
	Suppose it contains a vertex~$x \in V(G) \setminus (N^{\lceil d/2\rceil}_G[v]\cup N^{\lceil d/2\rceil}_G[w])$.
	Then~$\dist_G(v, x) + \dist_G(w, x) > 2 \lceil d/2 \rceil \ge d$, a contradiction to~$x$ being on a path from~$v$ to~$w$ of length at most~$d$.
	The claim follows.
\lqed
\end{proof}

\subsection{When every habitat must be complete (\texorpdfstring{$d=1$}{d=1})}

For~$d=1$,
the problem is solvable in linear time:
Check whether each habitat induces a clique.
If so,
check if the union of the cliques is small enough.

\begin{observation}
 \label{obs:1cgbp}
 \CgbpAcr[1]{} is solvable in linear time.
\end{observation}

\begin{proof}
  We employ the following algorithm:
  For each~$i\in\set{r}$,
  let~$G_i := G[V_i]$ and return~\no{} if~$G_i$ is not a clique.
  Finally,
  return~\yes{} if~$|\bigcup_{i=1}^r E(G_i)|\leq k$,
  and~\no{} otherwise.

  Clearly,
  if the algorithm returns \yes,
  then~$\calI$ is \yes-instance.
  Conversely,
  let~$\calI$ be a \yes-instance
  and let~$F'$ be a solution to~$\calI$.
  We know that for every~$i\in\set{r}$,
  and any two vertices~$v,w\in V_i$,
  edge~$\{v,w\}$ must be in~$F'$.
  It follows that
  $\bigcup_{i=1}^r E(G_i)\subseteq F'$.
  Thus,
  $|\bigcup_{i=1}^r E(G_i)|\leq |F'|\leq k$ 
  and the algorithm correctly returns~\yes.
  \lqed
\end{proof}

\subsection{When each part is just two steps away (\texorpdfstring{$d=2$}{d=2})}
\label{ssec:2cgbp}

For~$d=2$,
\CgbpAcr{}
becomes \NP-hard already on quite restrictive inputs.
It is however, as we show at the end of this section, still fixed-parameter tractable when parameterized by $k$.

\begin{proposition}
 \label{prop:2cgbp}
 \CgbpAcr[2]{} is \NP-complete,
 even
 if~$r=1$ and
 the input graph is bipartite and of diameter three.
\end{proposition}

\begin{construction}
 \label{constr:2cgbp}
 Let~$\calI=(G,k)$ with~$G=(V,E)$ be an instance of \prob{Vertex Cover},
 and assume without loss of generality that~$V=\set{n}$.
 Construct an instance of~\CgbpAcr[2]{}
 with graph~$G'=(V',E')$, 
 habitat~$V_1$,
 and integer~$k'\ceq 2|E|+k+3$ as follows
 (see~\cref{fig:2cgbp} for an illustration).
 \begin{figure}[t]
  \centering
  \begin{tikzpicture}
    \def\xr{1.1}
    \def\yr{0.7}
    \def\ys{1.5*\yr}
    \def\teps{0.125}
    \tikzstyle{xnode}=[circle,fill,scale=0.6,draw,color=\thecolor]
    \tikzstyle{xnodex}=[label={[xshift=-0.4*\xr cm]0:$\cdots$},color=\thecolor]
    \tikzstyle{xedge}=[thick,-,color=\thecolor]
    \tikzstyle{xnodef}=[circle,scale=0.6,draw,color=\thecolor,fill=white]
    
    \node (xp) at (0,1*\ys)[xnodef]{};
    \node (x) at (0*\xr,2*\ys)[xnode]{};
    
    \foreach \x/\y in {1/xnode,2/xnodex,3/xnode,4/xnodex,5/xnode,6/xnodex,7/xnode,8/xnodex,9/xnode}{
      \node (e\x) at (\x*\xr-5*\xr,0)[\y]{};
    }
    \draw[dotted,rounded corners,draw] ($(e1.south west)-(\teps,\teps)$) rectangle ($(e9.north east)+(\teps,\teps)$);
    \foreach \x/\y in {1/xnodef,2/xnodex,3/xnodef,4/xnodex,5/xnodef,6/xnodex,7/xnodef}{
    \node (v\x) at (\x*\xr-4*\xr,-1*\ys)[\y]{};
    }
    \draw[dotted,rounded corners,draw] ($(v1.south west)-(\teps,\teps)$) rectangle ($(v7.north east)+(\teps,\teps)$);
    \node (z) at (0,-2*\ys)[xnode]{};

    \foreach \x/\y in {3/3,3/5,5/3,5/2,7/5,7/6}{
    \draw[xedge] (e\x) to (v\y);
    }
    \foreach \x/\y in {2/1,8/7}{
    \draw[draw=none] (e\x) to node[midway]{$\vdots$}(v\y);
    }

    \foreach \x in {1,...,7}{
    \draw[xedge] (z) to (v\x);
    }
    
    \foreach \x in {1,...,9}{
    \draw[xedge] (e\x) to (xp);
    }
    \draw[xedge] (xp) to (x);
    
    \node[right =of e9](y)[xnodef]{};
    \draw[xedge] (z) to [out=0,in=-90](y);
    \draw[xedge] (y) to [out=90,in=0](x);
    
    \newcommand{\dlabel}[3]{
      \node at (#1)[label={[align=center,font=\footnotesize,color=black]#2},
      label={[align=center,font=\scriptsize,color=black]-90:#3}]{};
    }
    \newcommand{\ltb}[3]{
      \contourlength{0.09em}
      \node at (#1)[label={[align=center,font=\footnotesize,color=black]\contour*{white}{#2}},label={[align=center,font=\scriptsize,color=black]-90:\contour*{white}{#3}}]{};
    }

    \ltb{xp}{$y'$}{}
    \ltb{x}{$y$}{$\in V_1$}
    
    \ltb{e3}{$e=\{i,j\}$}{$\in V_1$};
    \ltb{e5}{$e'=\{i,j'\}$}{$\in V_1$};
    \ltb{e7}{$e''=\{i',j\}$}{$\in V_1$};

    \ltb{v3}{$v_i$}{};
    \ltb{v5}{$v_j$}{};
    
    \ltb{z}{$x$}{$\in V_1$};
    \dlabel{y}{180:$z$}{};
    
  \end{tikzpicture}
  \caption{Illustration to~\cref{constr:2cgbp} for~\CgbpAcr[2]{}.}
  \label{fig:2cgbp}
 \end{figure}

 To construct~$G'$ and~$V_1$,
 add the vertex set~$V_E\ceq \{v_e\mid e\in E\}$ and add~$V_E$ to~$V_1$.
 Add two designated vertices~$y'$ and~$y$,
 add~$y$ to~$V_1$,
 and make~$y'$ adjacent with~$y$ and all vertices in~$V_E$.
 Add a designated vertex~$x$,
 add~$x$ to~$V_1$,
 and introduce a path of length two from~$x$ to~$y$ (call the inner vertex~$z$).
 Add the vertex set~$V_G\ceq \{v_i\mid i\in V\}$,
 and make each~$v_i$ adjacent with~$x$ 
 and all edge-vertices corresponding to an edge incident with~$i$,
 i.e.,
 add the edge set~$E_G\ceq \bigcup_{i\in V}\{\{v_i,v_e\}\mid i\in e\}$.
 \cqed
\end{construction}

\begin{observation}
 \label{obs:2cgbp}
 Let~$\calI' = (G', \{V_1\}, k')$ be an instance obtained from application of \cref{constr:2cgbp} on an instance~$\calI=(G, k)$ of \prob{Vertex Cover}.
 If~$\calI'$ is a \yes-instance,
 then for every solution~$F\subseteq E(G')$ 
 we have~$\{\{y,y'\},\{y,z\},\{z,x\}\}\cup \{\{y',v_e\}\mid e\in E(G)\}\subseteq F$.
\end{observation}

\begin{lemma}
 \label{lem:2cgbp:edinc}
 Let~$\calI=(G,k)$ be an instance of \prob{Vertex Cover}.
 Consider the instance~$\calI'=(G',\{V_1\},k')$ obtained from~$\calI$ using~\cref{constr:2cgbp}.
 If~$\calI'$ is a \yes-instance,
 then there is a solution~$F\subseteq E(G')$ such that~$|N_{G'[F]}(v_e)\cap V_G|=1$ for all~$e\in E(G)$.
\end{lemma}

\begin{proof}
 Note that in every solution,
 clearly we have~$|N_{G'[F]}(v_e)\cap V_G|\geq 1$.
 Suppose there is a minimal solution~$F$ such that there is at least one~$e=\{i,j\}\in E$ such that~$|N_{G'[F]}(v_e)\cap V_G|=2$.
 Let~$F$ be a solution with a minimum number of edges incident to vertices in~$V_E$.
 
 Since~$\dist_{G'[F]}(v_e,x)= 2$,
 at least one of the edges~$\{v_i,x_i\}$ or~$\{v_j,x_j\}$ are in~$F$.
 If both are present
 then we can remove one of the edges~$\{v_e,v_i\}$ or~$\{v_e,v_j\}$ incident with~$v_e$ to obtain a solution of smaller size.
 This yields a contradiction.
 
 Otherwise,
 assume there is exactly one edge, say~$\{v_e,v_i\}$, 
 contained in~$F$.
 Then exchanging~$\{v_e,v_j\}$ with~$\{v_j,x\}$ yields a solution with a lower number of edges incident to vertices in~$V_E$.
 A contradiction.
 \lqed
\end{proof}

\begin{lemma}
 \label{lem:2cgbp:cor}
 Let~$\calI=(G,k)$ be an instance of \prob{Vertex Cover}.
 Consider the instance~$\calI'=(G',\{V_1\},k')$ obtained from~$\calI$ using~\cref{constr:2cgbp}.
 Then~$\calI$ is a \yes-instance if and only if~$\calI'$ is a \yes-instance.
\end{lemma}

\begin{proof}
 \RD{}
 Let~$W\subseteq V$ be a vertex cover of size at most~$k$ in~$G$.
 We construct a solution~$F\subseteq E'$ as follows.
 Let~$F'$ denote the set of all edges required due to~\cref{obs:2cgbp}.
 Let~$F_V\ceq \{\{v_i,x\}\mid i\in W\}$.
 We define the auxiliary function~$g\colon E\to V'$ with~$g(\{i,j\})=v_{\min(\{i,j\}\cap W)}$.
 Let~$F_E\ceq \bigcup_{e=\{i,j\}\in E} \{v_e,g(e)\}$.
 Let~$F\ceq F'\cup F_V\cup F_E$. 
 Note that~$|F|=|F'|+|F_V|+|F_E|\leq |E|+3+|E|+k = k'$.
 Moreover, 
 every~$v_e\in V'$ is connected to~$x$ via a path~$(v_e,v_i,z)$, 
 for some~$i\in (e\cap W)$,
 of length two.
 Thus all vertex pairs in~$V_1$ are at distance at most two.
 
 \LD{}
 Let~$\calI'$ be a \yes-instance.
 Due to~\cref{lem:2cgbp:edinc},
 there
 is a solution~$F\subseteq E'$ such that~$\deg_{G'[F]}(v_e)=1$ for all~$e\in E$.
 Let~$W\ceq \{i\in V\mid \{v_i,x\}\in F\}$.
 We claim that~$W$ is a vertex cover.
 Suppose not,
 that is,
 there is an edge~$e\in E$ such that~$e\cap W=\emptyset$.
 That means that the unique neighbor of~$v_e$,
 say~$v_i$,
 is not adjacent with~$x$ in~$G'[F]$.
 Then,
 $v_e$ is not connected with~$x$ in~$G'[F]^2$,
 and hence~$F$ is no solution,
 a contradiction.
 \lqed
\end{proof}

We next show fixed-parameter tractability when parameterizing by $k$.
All the reduction rules that worked for~\RgbpAcr[2]{}
also work for~\CgbpAcr[2]{}.
It thus follows that~\CgbpAcr[2]{} admits a problem kernel of size exponentially in~$k$.
As with~\RgbpAcr[2]{},
the problem kernel presumably cannot be much improved.
This can be shown by combining the constructions of~\cref{prop:rgbp:nopk,prop:2cgbp}. 

\begin{corollary}
 \CgbpAcr[2]{} admits a problem kernel of size exponentially in~$k$
 and,
 unless~\NPincoNPslashpoly,
 none of size polynomial in~$k$,
 even if~$r=1$.
\end{corollary}

\subsection{When reaching each part is a voyage (\texorpdfstring{$d\ge3$}{d≥3})}
\label{ssec:3cgbp}

For~$d\geq 3$,
the problem
is
\W{1}-hard regarding the number~$k$ of green bridges,
even for one habitat.
The reduction is similar to the one for \cref{prop:3rgbp}.

\begin{proposition}
 \label{prop:3cgbp}
 \CgbpAcr{} with~$d\geq 3$ is \NP-complete and~\W{1}-hard when parameterized by the number~$k$,
 even if $r=1$.
\end{proposition}

\begin{proof}
	Let~$\I=(G)$ with~$G=(U^1,\dots,U^k,E)$ be an instance of \prob{Multicolored Clique}.
	Apply
	\cref{constr:3rgbp}
	to obtain instance~$\I''=(G',\{V_1,\dots,V_{\binom{k}{2}}\},k')$
	(recall that~$k'=\frac{d-1}{2}k+\binom{k}{2}$).
	Let~$\I'=(G',\{V_1'\},k')$ with~$V_1'\ceq \bigcup_{i=1}^{\binom{k}{2}} V_i=\{v_1,\dots,v_k\}$ be the finally obtained instance of~\CgbpAcr{}.
	We claim that~$\I$ is a \yes-instance if and only if~$\I'$ is a \yes-instance.

	\RD{}
	Let~$C$ be a multicolored clique in~$G$.
	Let~$z_i\ceq V(C)\cap U^i$.
	We claim that~$F$,
	consisting of the edges of each shortest path from~$v_i$ to~$z_i$ and the edge set~$E(C)$,
	is a solution to~$\I'$.
	Note that~$|F|=k'$.
	Moreover,
	for any two~$i,j\in\set{k}$,
	we have that~$v_i$ and~$v_j$ are of distance~$2\frac{d-1}{2}+1=d$.
	Hence,
	$F$ is a solution.

	\LD{}
	Let~$F$ be a solution to~$\I$.
	Since~$F$ must contain a path from~$v_i$ to some~$z_i\in U^i$
	for every~$i\in\set{k}$,
	there are at most~$\binom{k}{2}$ edges left to connect.
	Let~$Z\ceq \{z_1,\dots,z_k\}$ be the vertices such that~$v_i$ is connected with~$z_i$ in~$G[F][U^i]$.
	As
	\begin{linenomath*}\[d\geq \dist_{G'[F]}(v_i,v_j) = \dist_{G'[F]}(v_i,z_i)+\dist_{G'[F]}(z_i,z_j)+\dist_{G'[F]}(z_j,v_j)\]\end{linenomath*}
	and $d-1=\dist_{G'[F]}(v_i,z_i)+\dist_{G'[F]}(z_j,v_j)$,
	it follows that~$\dist_{G'[F]}(z_i,z_j)=1$.
	Thus,
	$G[Z]$ forms a multicolored clique.
	\lqed
\end{proof}

\section{Connecting Habitats at Small Diameter}
\label{sec:dgbp}

Lastly,
we consider requiring short pairwise reachability
in \RgbpAcr[1]{}.

\decprob{\DgbpTsc{} (\DgbpAcr{})}{dgbp}
{An undirected graph~$G=(V,E)$,
a set~$\calH=\{V_1,\dots,V_r\}$ of habitats where~$V_i\subseteq V$ for all~$i\in\set{r}$, 
and an integer~$k\in\Nzero$.}
{Is there a subset~$F\subseteq E$ with~$|F|\leq k$ such that
for every~$i\in\set{r}$
it holds that~$V_i\subseteq V(G[F])$ and~$G[F][V_i]$ has diameter~$d$?
}

In particular,
$G[F][V_i]$ is required to be connected.
Note that
\RgbpAcr[1]{} reduces to~\prob{Diam \GBP} 
(where $d$ is part of the input and then set to the number of vertices in the input instance's graph).
We have the following.

\begin{theorem}
 \label{thm:dgbp}
 \DgbpAcr{} is,
 \begin{compactenum}[(i)]
  \item if~$d=1$, solvable in linear time;\label{thm:dgbp:i}
  \item if~$d=2$, \NP-hard even if~$r=1$.
 \end{compactenum}
 Moreover,
 \DgbpAcr{} admits a problem kernel with at most~$2k$ vertices and at most~$2^{2k}$ habitats.
\end{theorem}

\DgbpAcr[1]{} is equivalent to \CgbpAcr[1]{},
which is linear-time solvable as observed in \cref{obs:1cgbp}.
Thus, 
\cref{thm:dgbp}\eqref{thm:dgbp:i} follows.
Applying \cref{rr:few-habitat-vertices,rr:habitats} and deleting all non-habitat vertices yields the problem kernel.
At the end of this section we show that \DgbpAcr[2]{} most likely does not admit a polynomial kernel with respect to~$k$.
We now show that \DgbpAcr[2]{} is \NP-hard even if there is only one habitat.

\begin{proposition}
 \label{prop:2dgbp}
 \DgbpAcr[2]{} is \NP-hard even if~$r=1$.
\end{proposition}

\begin{construction}
  \label{constr:2dgbp}
  Let~$\I=(G,k)$ with~$G=(V,E)$ be an instance of~\prob{Vertex Cover}
  and assume without loss of generality that~$V=\{1,\dots,n\}$ and~$E=\{e_1,\dots,e_m\}$.
  Construct an instance~$\I'\ceq (G',\{V_1\},k')$
  with~$k'\ceq3m+2n+12+k$ as follows
  (see~\cref{fig:2dgbp} for an illustration).
\begin{figure}[t]
  \centering
  \begin{tikzpicture}

    \def\xr{1.}
    \def\yr{1.3}
    \def\ys{1.5*\yr}
    \def\teps{0.125}
    \tikzstyle{xnode}=[circle,fill,scale=0.6,draw,color=\thecolor]
    \tikzstyle{xnodef}=[circle,scale=0.6,draw,color=\thecolor,fill=white]
    \tikzstyle{xnodex}=[label={[xshift=-0.4*\xr cm]0:$\cdots$},color=\thecolor]
    \tikzstyle{xedge}=[thick,-,color=\thecolor]

    \foreach \x/\y in {1/xnode,2/xnodex,3/xnode,4/xnodex,5/xnode,6/xnodex,7/xnode,8/xnodex,9/xnode}{
    \node (e\x) at (\x*\xr-5*\xr,0)[\y]{};
    }
    \draw[dotted,rounded corners,draw] ($(e1.south west)-(\teps,\teps)$) rectangle ($(e9.north east)+(\teps,\teps)$);
    \foreach \x/\y in {1/xnodef,2/xnodex,3/xnodef,4/xnodex,5/xnodef,6/xnodex,7/xnodef}{
    \node (v\x) at (\x*\xr-4*\xr,-1*\ys)[\y]{};
    }
    \draw[dotted,rounded corners,draw] ($(v1.south west)-(\teps,\teps)$) rectangle ($(v7.north east)+(\teps,\teps)$);
    \node (x) at (0*\xr,-2*\ys)[xnodef]{};

    \node (y1) at (5*\xr,-0.5*\ys)[xnodef]{};
    \node (y2) at (5*\xr,-1*\ys)[xnodef]{};
    \node (y3) at (5*\xr,-1.5*\ys)[xnodef]{};
    
    \node (z1) at (5.5*\xr,-0.875*\ys)[xnodef]{};
    \node (z2) at (6*\xr,-0.75*\ys)[xnodef]{};
    \node (z3) at (6*\xr,-1.25*\ys)[xnodef]{};
    \node (z4) at (5.5*\xr,-1.125*\ys)[xnodef]{};

    \foreach \x/\y in {3/3,3/5,5/3,5/2,7/5,7/6}{
      \draw[xedge] (e\x) to (v\y);
    }
    \foreach \x in {1,...,9}{
        \draw[xedge] (e\x) to [out=-15,in=175](y1);
    }
    \foreach \x in {1,...,9}{
        \draw[xedge] (e\x) to [out=15,in=135](z3|-e1) to [out=-45,in=0](z3);
    }
    \foreach \x/\y in {2/1,8/7}{
      \draw[draw=none] (e\x) to node[midway]{$\vdots$}(v\y);
    }
    \foreach \x in {1,...,7}{
        \draw[xedge] (v\x) to [out=15,in=185](y1);
        \draw[xedge] (v\x) to [out=-15,in=180](y3);
    }
    \foreach \x in {1,...,7}{
      \draw[xedge] (x) to (v\x);
    }
    
    \draw[xedge] (y1) to (y2) to (y3);
    \draw[xedge] (z1) to (y1) to (z2) to (z3) to (y3) to (z4);
    \draw[xedge] (z1) to (z2) to (z3) to (z4) to (z1);

    \draw[xedge] (x) to (y3);
    \draw[xedge] (x) to [out=-15,in=-135](z2|-x) to [out=45,in=0](z2);
    
    \newcommandx{\ltb}[5][2=90,4=-90]{
      \contourlength{0.09em}
      \node at (#1)[inner sep=1pt,label={[align=center,font=\footnotesize,color=black]#2:\contour*{white}{#3}},label={[align=center,font=\scriptsize,color=black]#4:\contour*{white}{#5}}]{};
    }
    
    \ltb{e3}{$e=\{i,j\}$}{};
    \ltb{e5}{$e'=\{i,j'\}$}{};
    \ltb{e7}{$e''=\{i',j\}$}{};
    
    \ltb{x}{$x$}{};
    \ltb{y1}{$y_1$}{};
    \ltb{y2}{$y_2$}{};
    \ltb{y3}{}{$y_3$};
    \ltb{z1}{$z_1$}{};
    \ltb{z2}{$z_2$}{};
    \ltb{z3}{}{$z_3$};
    \ltb{z4}{}{$z_4$};
  
    \ltb{v3}{$v_i$}{};
    \ltb{v5}{$v_j$}{};
  \end{tikzpicture}
  \caption{Illustration for~\DgbpAcr[2]{} with~$r=1$.}
  \label{fig:2dgbp}
\end{figure}
  Add the vertex sets~$V_E\ceq \{v_e\mid e\in E\}$ and~$V_G=\{v_i\mid i\in V\}$,
  as well as the vertex set~$V_A\ceq \{x\}\cup\{y_i\mid i\in\{1,2,3\}\}\cup\{z_i\mid i\in\{1,\dots,4\}\}$.
  Add all vertices to~$V_1$.
  Next,
  for each~$e=\{i,j\}\in E$,
  connect~$v_e$ with~$v_i$, 
  $v_j$,
  $y_1$,
  and~$z_3$.
  For each~$i\in V$,
  connect~$v_i$ with~$x$, 
  $y_1$, 
  and~$y_3$.
  Lastly, 
  add the edge set~
  \begin{linenomath*}
  \begin{align*}
    E^*&\ceq\big\{\{y_1,y_2\},\{y_2,y_3\},\{y_1,z_1\},\{y_1,z_2\},\{y_3,z_3\},\{y_3,z_4\},\{y_3,x\},
    \\
    &\qquad \{z_1,z_4\},\{z_1,z_2\},\{z_2,z_3\},\{z_2,x\},\{z_3,z_4\}\big\}
  \end{align*}
  \end{linenomath*}
  to~$E'$.
  Let~$E_V^1 \ceq \{\{y_1,v_i\} \mid i \in V\}$,
  $E_V^3 \ceq \{\{y_3,v_i\} \mid i \in V\}$,
  $E_{E}^1\ceq\{\{y_1,v_e\} \mid e \in E\}$,
  and~$E_{E}^3\ceq\{\{z_3,v_e\} \mid e \in E\}$.
  \cqed
\end{construction}

\begin{observation}
 \label{obs:2dgbp}
 Let~$\I'$ be the instance obtained from applying \cref{constr:2dgbp} to some instance~$\I$.
 If~$\I'$ is a \yes-instance,
 then every solution~$F$ for~$\I'$ contains
 the edge set~$F'\ceq E^*\cup E_V^1\cup E_V^3\cup E_E^1\cup E_E^3$.
\end{observation}

\begin{proof}
 Let~$\I'$ be a \yes-instance and
 let~$F$ be a solution.
 Note that in~$G'-\{y_1\}$,
 there is no path of length at most two from any vertex in~$V_E\cup V_G$ to~$z_1$.
 Hence,
 $E_V^1\cup E_E^1\subseteq F$.
 In~$G'-\{y_3\}$,
 there is no path of length at most two from any vertex in~$V_G\cup\{x\}$ to~$z_4$.
 Hence,
 $E_V^3\subseteq F$.
 In~$G'-\{z_3\}$,
 there is no path of length at most two from any vertex in~$V_E$ to~$z_4$.
 Hence,
 $E_E^3\subseteq F$.
 In~$G'-\{z_2\}$,
 there is no path of length at most two from~$x$ to~$z_1$.
 Lastly,
 it is not difficult to see that every edge in~$E^*$ must be in~$F$.
 \lqed
\end{proof}

We are set to prove the correctness of~\cref{constr:2dgbp}.

\begin{lemma}
 \label{lem:2dgbp}
 Let~$\I'$ be the instance obtained from applying \cref{constr:2dgbp} to some instance~$\I$.
 Then,
 $\I$ is a \yes-instance if and only if~$\I'$ is a \yes-instance.
\end{lemma}

\begin{proof}
 \RD{}
 Let~$S\subseteq V$ be a vertex cover of size~$k$.
 Let~$F'$ denote the set of all edges required to be in a solution due to \cref{obs:2dgbp}.
 Let~$F_V\ceq \{\{v_i,x\}\mid i\in S\}$.
 We define the auxiliary function~$g\colon E\to V_G$ with~$g(\{i,j\})=v_{\min(\{i,j\}\cap S)}$.
 Let~$F_E\ceq \bigcup_{e\in E} \{\{v_e,g(e)\}\}$.
 Let~$F\ceq F'\cup F_V\cup F_E$. 
 Note that~$|F|=|F'|+|F_V|+|F_E|\leq (2m+2n+12)+k+m = k'$.
 Next consider~$G'[F][V_1]$.
 Observe that~$\dist_{G'[F][V_1]}(v,w)\leq 2$ 
 for every vertices~$v\in V_G\cup V_E\cup V_A$ and~$w\in V_A\setminus \{x\}$,
 for every vertices~$v,w\in V_G$,
 for every vertices~$v,w\in V_E$,
 and for every vertices~$v\in V_G$ and~$w=\{x\}$.
 We claim that for all~$e \in E$, 
 $\dist_{G'[F][V_1]}(x,v_e)=2$.
 By construction,
 $\dist_{G'[F][V_1]}(x,v_e)>1$.
 Suppose that there is~$v_e$ with~$e=\{i,j\}$ and~$\dist_{G'[F][V_1]}(x,v_e)>2$.
 Then there is no path~$(x,v,v_e)$ with~$v\in\{v_i,v_j\}$.
 Then~$\{i,j\}\cap S=\emptyset$,
 contradicting the fact that~$S$ is a vertex cover.
 
 \LD{} 
 Let~$F$ be a solution to~$\I'$.
 Let~$F'$ be the set of edges mentioned in \cref{obs:2dgbp};
 so~$F' \subseteq F$.
 Note that $|F'| = 2m + 2n + 12$.
 Observe that in~$G'-V_G$,
 the distance of~$x$ to any~$v_e\in V_E$ is larger than two.
 Hence,
 for each~$v_e$,
 there is a path~$(v_e,v,x)$ in~$G'[F][V_1]$ with~$v\in V_G$.
 We claim that~$S\ceq \{i\in V\mid \{v_i,x\}\in F\}$ is a vertex cover for~$G$ of size at most~$k$.
 Suppose not,
 that is,
 there is an edge~$e=\{i,j\}$ with~$e\cap S=\emptyset$.
 This contradicts the fact that there is a path~$(v_e,v,x)$ in~$G'[F][V_1]$ with~$v\in V_G$.
 It remains to show that~$|S|\leq k$.
 As~$F$ contains an edge~$\{v_e,v\}$ with~$v \in V_G$ for every~$e \in E$,
 $|S| = |F\cap \{\{v_i,x\} \mid i\in V\}| \le k'-(|F'|+m) = k$,
 and the claim follows.
 \lqed
\end{proof}

Additionally,
we have the following kernelization lower bound for~\DgbpAcr[2]{}.

\begin{proposition}
 Unless~$\NPincoNPslashpoly$,
 \DgbpAcr[2]{} admits no problem kernel of size polynomial in~$k$.
\end{proposition}

\begin{construction}
 \label{constr:2dgbp-nopk}
 Let~$\I=(U,\calF,k)$ with~$U=\{u_1,\dots,u_n\}$ and~$\calF=\{F_1,\dots,F_m\}$ be an instance of~\prob{Hitting Set}.
  Construct an instance~$\I'\ceq (G',\calH,k')$
  with~$k'=n+\binom{n}{2}+k$ as follows
 (see~\cref{fig:2dgbp-noPK} for an illustration).
 \begin{figure}[t]
  \centering
  \begin{tikzpicture}

    \def\xr{1.25}
    \def\yr{0.8}
    \def\ys{1.5*\yr}
    \def\teps{0.125}
    \tikzstyle{xnode}=[circle,fill,scale=0.6,draw,color=\thecolor]
    \tikzstyle{xnodef}=[circle,scale=0.6,draw,color=\thecolor,fill=white]
    \tikzstyle{xnodex}=[label={[xshift=-0.64*\xr cm]0:$\cdots$},color=\thecolor]
    \tikzstyle{xedge}=[thick,-,color=\thecolor]

    \foreach \x/\y in {1/xnode,2/xnodex,3/xnode,4/xnodex,5/xnode,6/xnodex,7/xnode,8/xnodex,9/xnode}{
    \node (e\x) at (\x*\xr-5*\xr,0)[\y]{};
    }
    \draw[dotted,rounded corners,draw] ($(e1.south west)-(\teps,\teps)$) rectangle ($(e9.north east)+(\teps,\teps)$);
    \foreach \x/\y in {1/xnodef,2/xnodex,3/xnodef,4/xnodex,5/xnodef,6/xnodex,7/xnodef}{
    \node (v\x) at (\x*\xr-4*\xr,-1*\ys)[\y]{};
    }
    \draw[dotted,rounded corners,draw] ($(v1.south west)-(\teps,\teps)$) rectangle ($(v7.north east)+(\teps,\teps)$);
    \node (x) at (0*\xr,-2*\ys)[xnodef]{};

    \foreach \x/\y in {3/3,3/5,5/3,5/2,7/5,7/6}{
      \draw[xedge] (e\x) to (v\y);
    }
    \foreach \x/\y in {2/1,8/7}{
      \draw[draw=none] (e\x) to node[midway]{$\vdots$}(v\y);
    }
    \foreach \x in {1,...,7}{
      \draw[xedge] (x) to (v\x);
    }
    
    \contourlength{0.09em}
    \newcommand{\ltb}[3]{
      \node at (#1)[label={[align=center,font=\footnotesize,color=black]\contour*{white}{#2}},label={[align=center,font=\scriptsize,color=black]-90:\contour*{white}{#3}}]{};
    }
    
    \ltb{e3}{$F$}{$\in V_F$};
    \ltb{e5}{$F'$}{$\in V_{F'}$};
    \ltb{e7}{$F''$}{$\in V_{F''}$};
    
    \ltb{x}{$x$}{$\in \bigcap_{F\in \calF} V_F$};
  
    \ltb{v3}{$v_i$}{$\in V_F \cap V_{F'}\cap V^i$};
    \ltb{v5}{$v_j$}{$\in V_{F'} \cap V_{F''}\cap V^j$};
  \end{tikzpicture}
  \caption{Illustration for~\DgbpAcr[2]{}. Here,
  $V^q$ denotes all sets~$V_{\{q,\cdot\}}$.}
  \label{fig:2dgbp-noPK}  
  \end{figure}

  Let~$V\ceq V_\calF \cup V_U\cup \{x\}$,
  where~$V_F\ceq \{v_F\mid F\in\calF\}$
  and~$V_U\ceq \{v_i\mid u_i\in U\}$.
  Add the edge sets~$E'\ceq \{\{v_i,v_{F_j}\}\mid u_i\in F_j\}$,
  $E_U\ceq \{\{v_i,v_j\}\mid \{i,j\}\in\binom{n}{2}\}$,
  and~$E_x\ceq \{\{x,v_i\}\mid i\in\set{n}$.
  The habitats~$\calH=\calH_\calF\cup\calH_U$ are defined as follows.
  For each~$F\in \calF$,
  there is the habitat~$V_F\in\calH_\calF$ with~$V_F\ceq \{x,v_F\}\cup\{v_i\mid u_i\in F\}$.
  For each~$\{i,j\}\in\binom{n}{2}$,
  there is the habitat~$V_{\{i,j\}}\in\calH_U$ with~$V_{\{i,j\}}\ceq \{v_i,v_j\}$.
  Finally,
  let~$k'\ceq n+\binom{n}{2}+k$.
\end{construction}

Due to~$\calH_U$,
we know that every solution needs to contain every edge in the clique induced by~$V_U$.

\begin{observation}
 \label{obs:2dgbp-nopk}
 Let~$\I'$ be the instance obtained from applying \cref{constr:2dgbp-nopk} to some instance~$\I$.
 If~$\I'$ is a \yes-instance,
 then every solution~$F$ for~$\I'$ contains
 the edge set~$E_U$.
\end{observation}

\begin{lemma}
 \label{lem:2dgbp-nopk}
 Let~$\I'$ be the instance obtained from applying \cref{constr:2dgbp-nopk} to some instance~$\I$.
 Then,
 $\I$ is a \yes-instance if and only if~$\I'$ is a \yes-instance.
\end{lemma}

\begin{proof}
 \RD{}
 Let~$S\subseteq U$ be a hitting set of size~$k$.
 We define the auxiliary function~$g\colon \calF\to V_U$ with~$g(F)=v_{\min\{i\mid u_i\in S\cap F\}}$.
 Let~$X_\calF\ceq \bigcup_{F\in \calF} \{\{v_F,g(F)\}\}$.
 Then~$X=E_U\cup X_\calF\cup\{\{x,v_i\}\mid u_i\in S\}$
 is a solution, as for every~$F \in \calF$,
 $G[X][V_F]$ contains as a subgraph a star with center~$g(F)$ and leaves~$x$ and~$V_U \setminus\{ g(F) \}$, thus it is of diameter at most two.
 
 \LD{}
 Let~$X$ be a solution to~$I'$.
 Due to~\cref{obs:2dgbp-nopk},
 we know that~$E_U\subseteq X$.
 Moreover, 
 every vertex in~$V_\calF$ has a neighbor in~$V_U$.
 We claim that~$S\ceq \{u_i\mid \{x,v_i\}\in X\}$
 is a solution to~$I$.
 Suppose not.
 Then there exists a set~$F \in \calF$ with~$S\cap F = \emptyset$.
 As $\diam(G[X][V_F]) \le 2$, we have that the distance between~$v_F$ and~$x$ is at most two.
 But then~$X$ must contain both~$\{v_F, v_i\}$ and~$\{v_i, x\}$ for some~$i \in \{1, \dots, n\}$.
 But then, by construction of~$E'$, we have $u_i \in S \cap F$, a contradiction.
\end{proof}

\section{Conclusion, Discussion, and Outlook}

We modeled the problem of placing wildlife crossings
with three different problem families:
\RgbpAcr{},
\CgbpAcr{},
and~\DgbpAcr{}.
We studied the practically desired cases~$d=1$ and $d=2$, 
as well as the cases~$d\ge3$.
For all three problems,
we settled the classic as well as the parameterized complexity 
(regarding the number~$k$ of wildlife crossings and the number~$r$ of habitats).
All three problems become~\NP-hard already for~$d=2$,
and~\RgbpAcr{} even for~$d=1$,
in most of the cases on restricted input graphs and only few habitats.
However,
all three variants are fixed-parameter tractable regarding~$k$ in the case of~$d=2$,
whereas, for $d \ge 3$, \RgbpAcr{} and~\CgbpAcr{} turn out to be intractable (yet in~$\XP$) for this parameter.
Thus,
the less desired cases~$d\geq 3$ are also algorithmically rather impractical.
Moreover,
\CgbpAcr{} and \DgbpAcr{} are tractable if the number $r$ of habitats and the maximum degree $\Delta$ of the graph are small,
which is expected to be likely in real-world applications.

\paragraph{Discussion.}
We derived an intriguing interrelation of
connection requirements,
data quality, 
and computational and parameterized complexity.
While each problem admits its individual complexity fingerprint,
each of them depends highly on the value of~$d$,
the level of the respective connectivity constraint.
This value can reflect the quality of the given data,
since naturally we assume that habitats are connected.
The worse the data, 
the stronger are the relaxations according to the connectivity of habitats,
and thus the larger is the value of~$d$.
Our results show that having very small ($d=2$) data gaps already leads to the problems becoming \NP-hard,
and that even larger gaps ($d\geq 3$) yield \W{1}-hardness (when parameterized by~$k$).
Hence, knowledge about habitats, connections, and data quality
decide which problem models can be applied,
thus influencing the computation power required to determine an optimal placement of wildlife crossings.
For instance,
for larger networks,
we recommend to ensure data quality such that one of our proposed problems for~$d\leq 2$ becomes applicable.
This
in turn
emphasizes the importance of careful habitat recognition.

In our models,
we neglected that different positions possibly lead to \emph{different} costs of building bridges 
(i.e.,\ edge~costs).
This neglect is justified when differentiating between types of bridges (and thus their costs)
is not necessary 
(e.g., if the habitat's species share preferred types of green bridges,
and the underlying human-made transportation lines are homogeneous).
In other scenarios,
additionally considering these costs may be beneficial for decision-making.

\paragraph{Outlook and open problems.}
As for algorithmic questions to the established problems, there are a few immediate questions that are unanswered in our work.
While \RgbpAcr[1]{} is \NP-hard even if~$r \ge 7$ but polynomial-time solvable if~$r \le 2$,
its complexity for~$2 < r < 7$ remains open.
Note that we obtained an~$\O(rd)$-approximation for~\RgbpAcr{},
which possibly leaves room for improvement and does not directly transfer to the other two problem variants.
It may be attractive to find out whether the problems admit FPT approximation algorithms as well.
For~$d \le 2$, all our problems allow for problem kernels where the number of vertices only depends on~$k$,
but it is presumed impossible to have a polynomial dependence on $k$.
If however the underlying street network is planar, then the input graphs to our problems can be seen as their planar dual.
Therefore, it is likely that the input graphs are planar in real-world applications.
In a follow-up work~\cite{fluschnik2022placing} we studied \RgbpAcr[1]{} with habitats that induce cycles and planar input graphs
and analyzed the algorithms (among them the approximation algorithm from \cref{prop:rgbp-approx}) on real-world graphs with synthetic habitats.

We conclude our work with some suggestions for extending our models.
Interesting directions here include,
for instance,
distinguishing types of green bridges to place,
taking into account possible movement directions within habitats (connectivity in directed graphs),
identifying real-world driven problem parameters leading to tractability,
or the problem of maintaining and servicing green bridges over time under a possible seasonal change of wildlife habitats 
(temporal graph modeling could fit well).

{
\begingroup
  \let\clearpage\relax
  \renewcommand{\url}[1]{\href{#1}{$\ExternalLink$}}
  \renewcommand{\doi}[1]{}
  \bibliography{green-bridges-bib}
\endgroup
}

\end{document}